\documentclass[a4paper,11pt]{article}
\pdfoutput=1
\usepackage{jcappub}
\usepackage{slashed}
\usepackage{bm} 
\usepackage{graphicx}
\usepackage{latexsym,amssymb,amsmath,float,url,mathrsfs}
\usepackage{array}
\usepackage{soul}
\usepackage{natbib}
\usepackage{hyperref}
\usepackage{slashed}
\usepackage{subfigure}
\usepackage[dvipsnames]{xcolor}

\hypersetup{
     colorlinks   = true,
     citecolor    = violet,
     urlcolor     = violet,
     linkcolor    = violet
}

\makeatletter
\newcommand{\pushright}[1]{\ifmeasuring@#1\else\omit\hfill$\displaystyle#1$\fi\ignorespaces}
\newcommand{\pushleft}[1]{\ifmeasuring@#1\else\omit$\displaystyle#1$\hfill\fi\ignorespaces}
\makeatother

\newcolumntype{M}[1]{>{\centering\arraybackslash}m{#1}}

\begin{document}

\title{Impact of Mass Generation for Simplified Dark Matter Models}

\author{Nicole F.\ Bell,}
\author{Yi Cai and}
\author{Rebecca K.\ Leane}
\affiliation{ARC Centre of Excellence for Particle Physics at the Terascale \\
School of Physics, The University of Melbourne, Victoria 3010, Australia}

\emailAdd{\tt n.bell@unimelb.edu.au}
\emailAdd{\tt yi.cai@unimelb.edu.au}
\emailAdd{\tt rleane@physics.unimelb.edu.au}

\date{\today}

\abstract{In the simplified dark matter models commonly studied, the
  mass generation mechanism for the dark fields is not typically
  specified. We demonstrate that the dark matter interaction types,
  and hence the annihilation processes relevant for relic density and
  indirect detection, are strongly dictated by the mass generation
  mechanism chosen for the dark sector particles, and the requirement
  of gauge invariance. We focus on the class of models in which
  fermionic dark matter couples to a spin-1 vector or axial-vector
  mediator.  However, in order to generate dark sector mass terms, it is
  necessary in most cases to introduce a dark Higgs field and thus a spin-0
  scalar mediator will also be present.  In the case that all the dark
  sector fields gain masses via coupling to a single dark sector Higgs
  field, it is mandatory that the axial-vector coupling of the spin-1
  mediator to the dark matter is non-zero; the vector coupling may
  also be present depending on the charge assignments.  For all other
  mass generation options, only pure vector couplings between the
  spin-1 mediator and the dark matter are allowed.  If these coupling
  restrictions are not obeyed, unphysical results may be obtained such
  as a violation of unitarity at high energies.  These two-mediator
  scenarios lead to important phenomenology that does not arise in
  single mediator models. We survey two-mediator dark matter models
  which contain both vector and scalar mediators, and explore their
  relic density and indirect detection phenomenology.}
\maketitle


\section{Introduction}
The search for particle interactions of dark matter (DM) is currently being pursued across a great variety of experiments.  Foremost amongst these are the searches for Weakly Interacting Massive Particles (WIMPs)~\cite{Bergstrom:2000pn,Bertone:2004pz}.  The WIMP mass and coupling parameters are being probed with unprecedented sensitivity at in direct detection experiments such as LUX~\cite{Akerib:2013tjd,Akerib:2015rjg} and PandaX-II~\cite{Tan:2016zwf}, the mono-$X$ searches at the Large Hadron Collider~\cite{Aaboud:2016tnv,Aaboud:2016uro,Aaboud:2016qgg,ATLAS:2016tsc,ATLAS:1437995,CMS:2016flr,CMS:2016fnh,CMS:2016hmx,CMS:2016pod,CMS:2016mjh,CMS:2016xok,CMS:2016mxc,CMS:2016uxr, Carpenter:2012rg, Carpenter:2013xra, Petrov:2013nia, Bell:2012rg, Bell:2015rdw, Birkedal:2004xn, Gershtein:2008bf, Goodman:2010ku, Crivellin:2015wva, Petriello:2008pu, Berlin:2014cfa, Lin:2013sca, Fox:2011pm, Bai:2015nfa, Autran:2015mfa, Gupta:2015lfa,Ghorbani:2016edw}, and in indirect detection analysis of 
astrophysical gamma-ray fluxes such as those measured by the Fermi-LAT satellite~\cite{Ackermann:2015zua}.  The physics reach of these searches is such that we can realistically expect to cover much of the WIMP parameter space in the near future.  As such, it is imperative to have well-formulated models of DM interactions which span a comprehensive spectrum of possible interaction types, in a manner which is as model independent as possible.  Simplified models address this aim by introducing a single DM candidate and a mediator which communicates between the dark and SM sectors ~\cite{Abdallah:2014hon,Buckley:2014fba,Alves:2011wf,Alwall:2008ag,Abdallah:2015ter,Abercrombie:2015wmb, DeSimone:2016fbz,Jacques:2016dqz}.  The three most commonly considered benchmark simplified models involve the interaction of fermionic DM with Standard Model (SM) fermions via a spin-1 $s$-channel mediator, a spin-0 $s$-channel mediator, or a spin-0 $t$-channel mediator~\cite{Abercrombie:2015wmb}.

These simplified models are an improvement over the effective field theory approach~\cite{Goodman:2010yf,Goodman:2010ku,Duch:2014xda} which was used for many recent collider and non-collider WIMP searches, yet suffers from unitarity issues when used outside the region of validity~\cite{Shoemaker:2011vi,Fox:2012ee,Busoni:2013lha,Buchmueller:2013dya,Busoni:2014sya,Endo:2014mja,Busoni:2014haa,Hedri:2014mua,Bell:2015sza,Baek:2015lna,Bell:2016obu}. However, the simplified models are still far from ideal.  Indeed, by their simplified nature, they are are not intrinsically capable of capturing the realistic phenomenology of many UV complete theories, which may have multiple dark-sector field content. More critically, separate consideration of the benchmark simplified models can lead to scenarios that are not physically viable. Indeed, the simplified models suffer some of the same issues that plague the effective field theory approach, such as violations of perturbative unitarity that arise because gauge invariance 
is not respected~\cite{Shoemaker:2011vi,Fox:2012ee,Busoni:2013lha,Buchmueller:2013dya,Busoni:2014sya,Endo:2014mja,Busoni:2014haa,Hedri:2014mua,Bell:2015sza,Baek:2015lna,Kahlhoefer:2015bea,Bell:2015rdw,Haisch:2016usn,Englert:2016joy,Bell:2016obu}.

As an example of such an issue, simplified models in which the DM has a non-zero axial-vector coupling to a spin-1 mediator will violate perturbative unitarity at high energies~\cite{Cline:2014dwa, Kahlhoefer:2015bea}.  This can be remedied by introducing a dark Higgs field to unitarize the longitudinal component of the $Z'$~\cite{Cline:2014dwa, Kahlhoefer:2015bea, Bell:2016fqf, Duerr:2016tmh}.  The dark Higgs may also provide mass to the DM itself.  The minimal self-consistent approach is then a multi-mediator model, featuring both spin-1 and spin-0 mediators\footnote{Multi-mediator models have also been considered recently in Refs.~\cite{Cline:2015qha,Choudhury:2015lha,Ghorbani:2015baa, Bell:2016fqf, Duerr:2016tmh}.}.  This of course can alter the phenomenology, even at low energies. In our recent work~\cite{Bell:2016fqf}, we considered indirect detection signals in a scenario with a Majorana DM candidate $\chi$, in which the couplings of a $Z'$ and scalar, $s$, are related by gauge invariance.  In 
this scenario, the presence of both the $s$ and $Z'$ mediators opens a dominant $s$-wave annihilation channel, $\chi\chi\rightarrow sZ'$, that does not arise when a single-mediator is considered in isolation~\cite{Bell:2016fqf}.  This has a dramatic impact on the indirect detection phenomenology.

An important consideration for DM models is the mass
generation mechanism for the dark sector fields.  Although commonly
left unspecified in the simplified model approach, with mass terms
simply added by hand, we shall argue that the mechanism of mass
generation has significant consequences that cannot be ignored.  For a
spin-1 mediator with only vector couplings, a standard procedure is to
appeal to the Stueckelberg mechanism to introduce a mass for the
vector boson.  However, this is valid only for a pure vector, with
vanishing axial-vector couplings to fermions.  This is a very specific
scenario, and there is no reason to assume it is correct.  In fact, the
Higgs mechanism is the only mass generation mechanism we know is
realized by nature, as confirmed by the recent experimental discovery
of the SM Higgs boson.  As such, it is well motivated to consider a
variety of scenarios where different dark sector fields acquire their
mass by various methods: the Stueckelberg mechanism, a dark Higgs
mechanism, or in cases where it is allowed, simply with a bare mass term.

We will show that the annihilation processes, and hence both the relic
density and indirect detection constraints, are strongly dictated by
the mass generation mechanisms. Interestingly, we will also show that
depending on the choice of mass generation mechanism, only particular
interactions types are allowed, as dictated by dark gauge invariance.
In most cases, only pure vector couplings of the spin-1 mediator to
fermionic DM are allowed.  Conversely, if a single dark Higgs
mechanism gives mass to all the dark sector fields, the axial-vector
coupling of the spin-1 mediator to the DM is required to be
non-zero.  Such restrictions do not map to the single-mediator
simplified models, despite being a compelling possibility (or in some
cases, a requirement). Again, this phenomenology is not accurately
captured by the single mediator simplified model framework.

The purpose of this paper is to undertake a more complete study of simplified models that contain both a scalar and vector mediator.  
In all cases, we will be sure to enforce gauge invariance with respect to the dark $U(1)_\chi$ interaction (dark gauge invariance), which is important to ensure physically well behaved cross sections. We will consider Dirac DM, which allows for a wider combination of coupling types, each with their own distinct phenomenology. Results for Majorana DM can be obtained in the limit of one of the scenarios we investigate in this paper. We focus, in particular, on hidden-sector type models~\cite{Pospelov:2007mp,Pospelov:2008jd,Pospelov:2008zw,Feng:2008mu,Feng:2008ya,
  Rothstein:2009pm,Mardon:2009gw,Mardon:2009rc,
  Meade:2009rb,Cheung:2010gj, Davoudiasl:2013jma,
  Berlin:2014pya,Liu:2014cma, Hardy:2014dea,
  Boehm:2014bia,McDermott:2014rqa,Chacko:2015noa, Elor:2015tva,
  Elor:2015bho, Abdullah:2014lla, Martin:2014sxa, Ko:2015ioa,
  Ko:2014gha, Kim:2016csm, Hooper:2012cw, Berlin:2015wwa}, where the DM annihilates directly to the mediators, which then decay to SM particles via small couplings between the dark and visible sectors. In section \ref{sec:models} we outline mass generation for spin-1 simplified models, and in section \ref{sec:scenI} we briefly discuss the standard assumption for mass generation in spin-1 models, before investigating three other compelling mass generation scenarios in sections \ref{sec:scenII}, \ref{sec:scenIII} and \ref{sec:scenIV}, detailing models, annihilation processes and relic density constraints. We present indirect detection constraints in section \ref{sec:indirectdet} and summarize our findings in section \ref{sec:summary}.

\section{Mass Generation for Spin-1 Simplified Models}
\label{sec:models}

The mass generation mechanism for fermionic DM in spin-1 simplified models is tightly correlated with the DM interaction type.
In the case that DM is Majorana, the $Z'$ can have only axial-vector couplings to the DM, as vector couplings of Majorana particles vanish.  
In the case where DM is Dirac, both vector and axial-vector couplings to the $Z'$ can simultaneously be present. For both DM types, 
the presence of an axial-vector coupling is significant, as it implies that  
\begin{enumerate}
\item
The DM mass must arise after symmetry breaking, as the $U(1)_\chi$ gauge symmetry prevents a bare mass term for $\chi$, and
\item
A $U(1)_\chi$ symmetry breaking mechanism is required to give the $Z'$ mass, in order to unitarize the longitudinal component of the $Z'$. 
\end{enumerate}
A single dark Higgs field is an economical solution to these issues.
In the following sections, we will show that the only scenario in which an
axial-vector coupling is possible in a spin-1 mediator model is if
there is a dark Higgs which interacts with both the DM and
the dark gauge boson.  Moreover, the axial coupling is not merely
possible in this case, but in fact required to be non-zero by gauge
invariance. We take the DM to be Dirac, as this permits the broadest
range of possible coupling types.  A related model involving Majorana
fermions can be found in Ref.~\cite{Bell:2016fqf} and is closely
related to a specific realization of scenario II presented below.

\begin{table}
\centering
\begin{tabular}{|c|M{2.0cm}|M{2.1cm}|M{3.7cm}|M{2.8cm}|M{0.6cm}|}
\hline
\textbf{Scenario} & \textbf{ $\chi$ mass} & \textbf{$Z'$ mass} & \textbf{Required $\chi-Z'$ coupling type}  & \textbf{ Annihilation processes} & \textbf{$Z'$ pol} \\
\hline\hline
\hyperref[sec:scenI]{I} & Bare mass term & Stueckelberg mechanism & Vector & 
\parbox[c]{3cm}{\includegraphics[width=2.9cm]{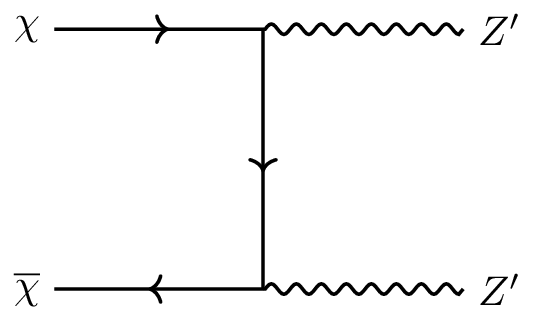}} & $Z'_T$
\\ \hline
\hyperref[sec:scenII]{II}  & Yukawa coupling to Dark Higgs & Dark Higgs mechanism & 
\begin{center}Vector $\&$ axial-vector\\or\\pure axial-vector.\end{center} 
The $U(1)$ charge assignments of $\chi_L$ and $\chi_R$ determine the relative size of the V and A couplings. \textbf{The axial-vector coupling must be non-zero}. &
\parbox[t][3.5cm][c]{3cm}{\includegraphics[width=2.9cm]{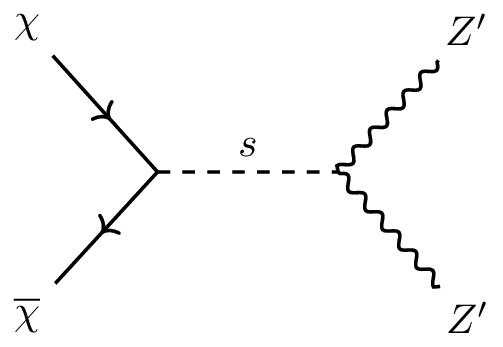} \newline\includegraphics[width=2.9cm]{tchan_zpzp.png}}
\parbox[t][4.0cm][c]{3cm}{\includegraphics[width=2.9cm]{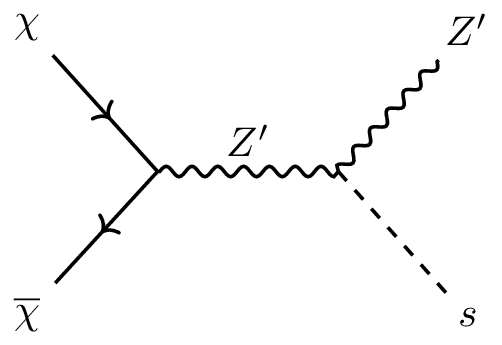}\newline\includegraphics[width=2.9cm]{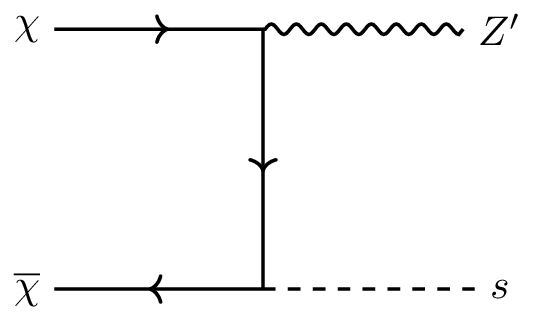}} & $Z'_T$ \& $Z'_L$
\\ \hline
\hyperref[sec:scenIII]{III} & Yukawa coupling to Dark Higgs & Stueckelberg mechanism & Vector & 
\parbox[c]{3cm}{\includegraphics[width=2.9cm]{tchan_zpzp.png}}
\parbox[c]{3cm}{\includegraphics[width=2.9cm]{tchan_zps.png}} & $Z'_T$
\\ \hline
\hyperref[sec:scenIV]{IV}  & Bare mass term & Dark Higgs mechanism & Vector & 
\parbox[c]{3cm}{\includegraphics[width=2.9cm]{tchan_zpzp.png}}
\parbox[c]{3cm}{\includegraphics[width=2.9cm]{schan_zps.png}} & $Z'_T$
\\ \hline
\end{tabular}
\caption{The spectrum of scenarios with distinct phenomenology once mass generation is specified. All $t$-channel annihilation processes have an accompanying $u$-channel process which is not shown. All processes shown are $s$-wave, except for $\chi\overline{\chi}\to s \to Z'Z'$ diagram in scenario II, which while $p$-wave when considered alone, is part of the process $\chi\overline{\chi}\to Z'Z'$. For scenario III, as the dark Yukawa and gauge coupling are not correlated, the $p$-wave annihilation to two dark Higgs, $\chi\overline{\chi}\rightarrow ss$, can have an impact on the relic density if the gauge coupling is sufficiently small to suppress the $s$-wave processes. Otherwise, the $s$-wave processes shown dominate, even at freeze-out. The final column displays the polarization of the $Z'$ bosons produced by these annihilation processes (in the $E^2_{Z'}\gg m^2_{Z'}$ limit).}
\label{table:cases}
\end{table}

For Dirac DM, it is possible to have pure vectorlike couplings to the
$Z'$ and so it is possible to include a bare mass term for DM, and use
the Stueckelberg mechanism\footnote{In Abelian gauge theories, the
  Stueckelberg mechanism can be taken as the limit of the Higgs
  mechanism where the mass of the real scalar is sent to infinity and
  only the pseudoscalar is present; however it is not always easily
  realized in more complicated scenarios.  In particular, unitarity is
  already violated at tree-level in a non-Abelian theory with a
  Stueckelberg Lagrangian and thus the theory is not
  renormalizable~\cite{Kunimasa:1967zza,Slavnov:1972fg,
    Veltman:1968ki,Slavnov:1970tk,Vainshtein:1971ip,Shizuya:1976qf,McKeon:1991hv,Freedman:1980us,Kafiev:1982hx}.
  In general the Stueckelberg mechanism should be treated as an
  alternative to the Higgs mechanism for mass generation.}  to provide
a mass for the $Z'$, such that no dark Higgs is needed.  Nonetheless,
even in the case of pure vector couplings, a dark Higgs may still
provide mass for one or both of the $Z'$ and DM.  Furthermore, when
the $Z'$ and DM masses arise from different mechanisms, the coupling
of the DM to the scalar and vector mediators are no longer related to
each other, and hence the phenomenology is less constrained.  We are
thus led to a spectrum of models in which both scalar and vector
mediators would be present. We outline the phenomenologically distinct
scenarios in Tab.~(\ref{table:cases}).

\section{Scenario I: Bare DM Mass and $Z'$ Mass from Stueckelberg Mechanism}
\label{sec:scenI}
\vspace{2mm}
\centerline{\textit{Interaction type required: Pure Vector}}
\vspace{4mm}

This is the most minimal gauge-invariant scenario, and is permitted
only if there are pure vectors couplings between the DM and the
$Z'$. 
Unlike the axial-vector scenario, a dark Higgs is not mandatory in the
pure vector case because
\begin{enumerate}
\item
The $Z'$ gauge boson can acquire a mass via the Stueckelberg mechanism.
\item
As $\chi$ is vectorlike with respect to the $U(1)_\chi$,
i.e. $Q_{\chi_R}=Q_{\chi_L}$, a bare $\chi$ mass term is permitted.
\end{enumerate}

\subsection{Model}
The Lagrangian for this case is simply
\begin{align}
 \mathcal{L}= \mathcal{L}_{SM}
+ i\, \overline{\chi} (\partial_\mu +ig_\chi Q_V Z'_\mu)\gamma^\mu \chi  
-\frac{\sin\epsilon}{2} Z'^{\mu\nu}B_{\mu\nu}-m_\chi\overline{\chi}\chi +\frac{1}{2}m_{Z'}^2  Z'^\mu Z'_\mu.
\end{align}
where $Q_V$ is the vectorlike $U(1)_\chi$ charge of the DM, which can
be chosen freely, and the $\epsilon$ term describes kinetic mixing of the
$U(1)_\chi$ gauge boson with the SM hypercharge gauge boson.  This is
the only spin-1 mediator scenario where it is possible to avoid the
inclusion of a dark Higgs. This case has been thoroughly covered in
the literature (for a review see, e.g., \cite{DeSimone:2016fbz}); we do not
discuss it further.
 
 \section{Scenario II: DM Mass and $Z'$ Mass both from Dark Higgs Mechanism}
\label{sec:scenII}

\vspace{2mm}
\centerline{\textit{Interaction type required: Non-Zero Axial-Vector}}
\vspace{4mm}

We now consider the case where both the DM and the dark gauge
boson acquire mass from a single dark Higgs.  We will show that this
requires the axial-vector DM-$Z'$ interaction to be non-zero.  The
reason is simple: the dark Higgs field, $S$, must clearly carry
$U(1)_\chi$ charge if its vacuum expectation
value (vev) is to break that symmetry.  A Yukawa
coupling of the dark Higgs to the DM of the form $y_\chi \chi_R \chi_L
S$ is then possible only if the DM is chiral, i.e. $\chi_L$ and
$\chi_R$ carry different $U(1)_\chi$ charges.  This guarantees that
the axial couping is non-zero (while the vector couplings may be
either zero or non-zero depending on the $U(1)_\chi$ charge
assignments).

\subsection{Model}

We investigate the phenomenology of the most minimal model containing
a dark gauge boson and a dark Higgs field, by simply extending the
Standard Model by an extra $U(1)$.  The gauge group is thus:
$SU(3)_c\otimes SU(2)_W\otimes U(1)_Y\otimes U(1)_\chi$.  Here the
covariant derivative is $D_\mu=D_\mu^{SM}+i Q g_{\chi}Z'_\mu$,
where $Q$ denotes the $U(1)_\chi$ charge. The SM field content
is augmented by a Dirac fermion DM candidate, $\chi$, a spin-1 dark
gauge boson, $Z'$, and a dark Higgs field $S$. The vev of the dark Higgs field provides a mass generation
mechanism for the dark sector fields $Z'$ and $\chi$.  Before
electroweak and $U(1)_\chi$ symmetry breaking, the most general
Lagrangian is
\begin{align}
\mathcal{L} = \mathcal{L}_{\rm SM} 
&+ i \overline{\chi}_L\slashed{D}\chi_L 
+ i \overline{\chi}_R\slashed{D}\chi_R 
- \left(y_\chi \overline{\chi}_R {\chi}_L S + h.c.\right) \nonumber\\
& + (D^\mu S)^\dagger (D_\mu S)
- \mu_s^2 S^\dagger S - \lambda_s (S^\dagger S)^2  - \lambda_{hs}(S^\dagger S)(H^\dagger H) 
-\frac{\sin\epsilon}{2} Z'^{\mu\nu}B_{\mu\nu}.
\end{align}
We assume that the SM fields are not charged under $U(1)_\chi$.  There
are thus only two possible terms that couple SM and dark-sector
fields: the kinetic mixing of the $U(1)$ gauge boson with the
hypercharge gauge boson, controlled by the kinetic mixing parameter
$\epsilon$, and mixing of the dark Higgs, $S$, with the SM Higgs, $H$,
controlled by the Higgs mixing parameter $\lambda_{hs}$.

In order for the $\chi$-$S$ Yukawa term to be gauge invariant, the
charges of the dark sector field must be chosen to
satisfy\footnote{For anomaly cancellation there must be additional
  fields charged under the dark $U(1)$. However, we include only the
  lightest of such fields as the DM candidate, as the others can be
  made heavier such that the phenomenology is not affected, as they
  make a subdominant contribution to the relic density, as can be seen
  from section \ref{sec:relicdensII}.}
\begin{align}
Q_{\chi_R} - Q_{\chi_L} = Q_S \; . 
\end{align}   
We can set the dark Higgs charge to be $Q_S=1$, without loss of
generality, as any other choice can be absorbed into a rescaling of the
dark gauge coupling.  The $\chi$ charges therefore satisfy
\begin{align}
Q_A \equiv \frac{1}{2} (Q_{\chi_R} - Q_{\chi_L}) &= \frac{1}{2}, \\
Q_V \equiv \frac{1}{2} (Q_{\chi_R} + Q_{\chi_L}) &= \frac{1}{2} + Q_{\chi_L}. 
\end{align}
These charges determine the vector and axial-vector couplings of the
$Z'$ to the $\chi$. We see that the axial-vector coupling is
completely determined, while there is freedom to adjust the vector
coupling by choosing $Q_{\chi_{L,R}}$ appropriately.  For instance,
$Q_{\chi_L} = 0$ would lead to equal vector and axial-vector
couplings, while $Q_{\chi_L} \gg 1$ would lead to a vector coupling
much larger than the axial-vector.  Pure axial-vector is obtained with
$Q_{\chi_L} = - 1/2$; this produces phenomenology similar to the
Majorana model studied in \cite{Bell:2016fqf}.  Pure vector, on the
other hand, can only be approximately reached in the limit $Q_V \gg
1$, but never fully realized\footnote{It is also important to note
  that there exist relations between the axial-vector coupling size
  and the masses of the dark sector fields
  \cite{Kahlhoefer:2015bea}. Therefore, for almost all mass choices of
  the dark sector fields, it is not possible to make the axial-vector
  coupling vanishingly small relative to the vector coupling without
  the vector coupling becoming non-perturbative. Thus, the
  axial-vector coupling is effectively never negligible and cannot be
  neglected even in limiting cases.}, as dark gauge invariance
prevents the axial-vector from being exactly zero.

Both $S$ and $H$ obtain vevs, breaking $SU(3)_c\otimes SU(2)_L\otimes
U(1)_Y\otimes U(1)_\chi$ down to $SU(3)_c\otimes U(1)_{em}$.  In the
broken phase, the terms of interest are
\begin{align}
 \mathcal{L}\supset & 
-\frac{1}{2}m_{s}^2 s^2 +\frac{1}{2}m_{Z'}^2  Z'^\mu Z'_\mu -m_\chi \overline{\chi}\chi 
\nonumber \\ 
& +  g_\chi^2 w Z'^\mu Z'_\mu s - \lambda_s w s^3 - 2\lambda_{hs} h \, s (v \, s+ w \, h) 
+ g_f \sum_f Z'_\mu \overline{f}\Gamma^\mu_f f \\ 
& -  g_\chi Q_V Z'_\mu \overline{\chi}\gamma^\mu \chi 
\nonumber
- g_\chi Q_A Z'_\mu \overline{\chi}\gamma^\mu \gamma_5 \chi 
  -\frac{y_\chi}{\sqrt{2}} s \overline{\chi}\chi \; , 
\label{eq:brokenlgn}
 \end{align}
where the component fields of $S$ and $H$ are defined in the broken
phase as $S\equiv \frac{1}{\sqrt{2}}(w + s + ia)$ and $H=\left\{ G^+,
\frac{1}{\sqrt{2}}(v + h + iG^0)\right\}$ with $G^+$, $G^0$ and $a$
being the Goldstone bosons of $W$, $Z$ and $Z^\prime$ respectively,
while $s$ and $h$ are real scalars. The coupling $g_f$, which
controls the interactions of the $Z'$ with SM fermions, is dictated by
the kinetic mixing; the explicit form can be found, e.g., in
Ref.~\cite{Agashe:2014kda}. We assume that the scalar mixing parameter
$\lambda_{hs}$ is small, which implies that the the SM Higgs is not
significantly perturbed by the new physics.  In this limit, the dark
Higgs vev satisfies $w^2=-\mu_s^2/\lambda_s$ and the various masses
are:
\begin{subequations}
\begin{align}
   m_{Z'}& =g_\chi w, \\
   m_{\chi}& =\frac{1}{\sqrt{2}}y_\chi w, \\
   m_{s}^2 & \simeq  - 2\mu_s^2, \\
   m_{h}^2 & \simeq -2 \mu_h^2 \; . 
\end{align}
 \label{eq:masses}
\end{subequations}
Importantly, because both the DM and $Z'$ masses are both proportional
the to vev of the dark Higgs, their masses and couplings are not all
independent parameters but instead are related as
\begin{equation}
\label{eq:coupling}
y_\chi/g_\chi = \sqrt{2} m_\chi/m_{Z'}.
\end{equation}

\subsection{Cross Sections}

The relevant annihilation process for this scenario are shown in
Tab.~(\ref{table:cases}).  The $\chi\overline{\chi} \rightarrow sZ'$
annihilation receives contributions from both $s$ and $t/u$ channel
processes, while $s$-wave contributions to the $\chi\overline{\chi}
\rightarrow Z'Z'$ process arise only from the $t/u$ channel diagrams.
(Note, however, that the contribution of the $s$-channel scalar
exchange diagram to the annihilation to $Z'Z'$ is necessary to
unitarize the cross section at high energy.  Without this
contribution, longitudinal $Z_L'$ contributions would lead to
unphysical high energy behavior of the $p$-wave term.)  The $s$-wave
contributions to the annihilation cross sections are given by
\begin{align}
 \langle \sigma v \rangle_{\chi\overline{\chi}\to Z'Z'} & = \frac{g_{\chi }^4 \left(1-\eta _{Z'}\right){}^{3/2} \left(
16Q_V^4 \eta _{Z'} + 8 Q_V^2 \left(4-3 \eta _{Z'}\right)+ \eta _{Z'}
\right)}{64 \pi  m_{\chi }^2 \left(\eta _{Z'}-2\right){}^2 \eta _{Z'}},
\end{align}
and
\begin{align}
 \langle \sigma v \rangle_{\chi\overline{\chi}\to sZ'} & 
  =\frac{g_{\chi }^4 
\sqrt{ \left(\eta _s - \eta _{Z'} -4 \right)^2 - 16 \eta_{Z'}}}{1024 \pi  m_{\chi }^2 \left(\eta _{Z'}-4\right)^2 \eta _{Z'}^2 \left(\eta _s +\eta _{Z'} - 4\right)^2} \times\nonumber\\
 & \Big\{  \left(\eta _{Z'}-4\right)^2 
(\eta_s + \eta_{Z'}-4)^2 \left( (\eta _s -\eta _{Z'} -4)^2  - 16 \eta_{Z'}\right)
\\
 &+4Q_V^2 \eta _{Z'} \Big[\eta _s^4 \eta _{Z'}-16 \eta _s^3 \eta _{Z'}-2 \eta _s^2 \left(\eta _{Z'}^3-44 \eta _{Z'}^2+80 \eta _{Z'}-64\right)  \nonumber\\
 &+ 64  \eta _s \left(\eta _{Z'}-4\right){}^2 \left(\eta _{Z'}-1\right)+ \left(\eta _{Z'}-4\right){}^4 \left(\eta _{Z'}+8\right)\Big] \Big\},\nonumber
\end{align}
where $\eta_{s,Z'}=m_{s,Z'}^2/m_\chi^2$.  As explained above, we have
set $Q_S = 1 = 2Q_A$ without loss of generality, while $Q_V$ is left as a free
parameter.  Also note that we have used Eq.~(\ref{eq:coupling}) to replace the Yukawa coupling $y_\chi$ with the gauge coupling $g_\chi$.  

\begin{figure}[h]
\begin{center}
        \subfigure{
            \includegraphics[width=0.32\columnwidth]{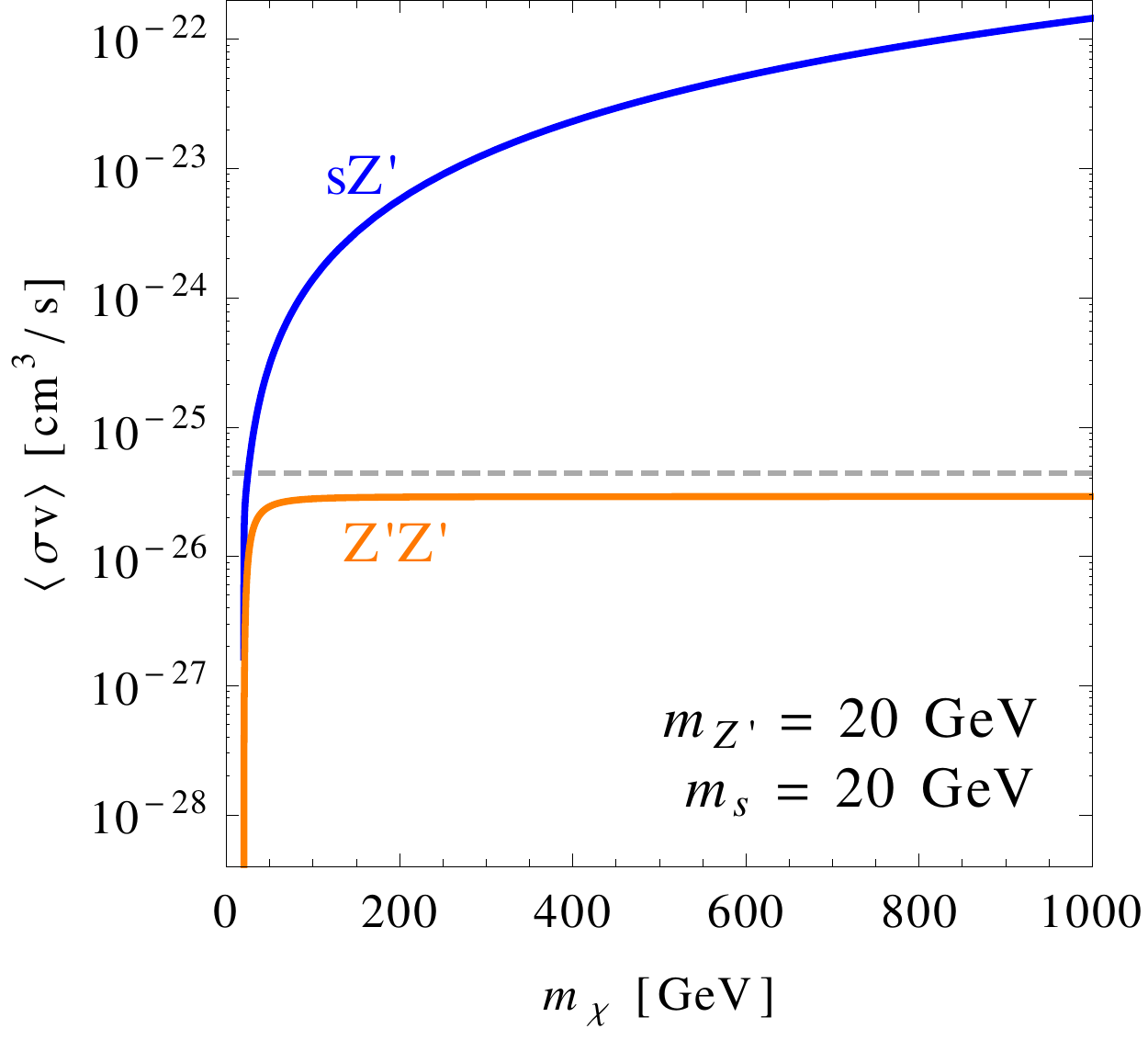}}
        \subfigure{
            \includegraphics[width=0.32\columnwidth]{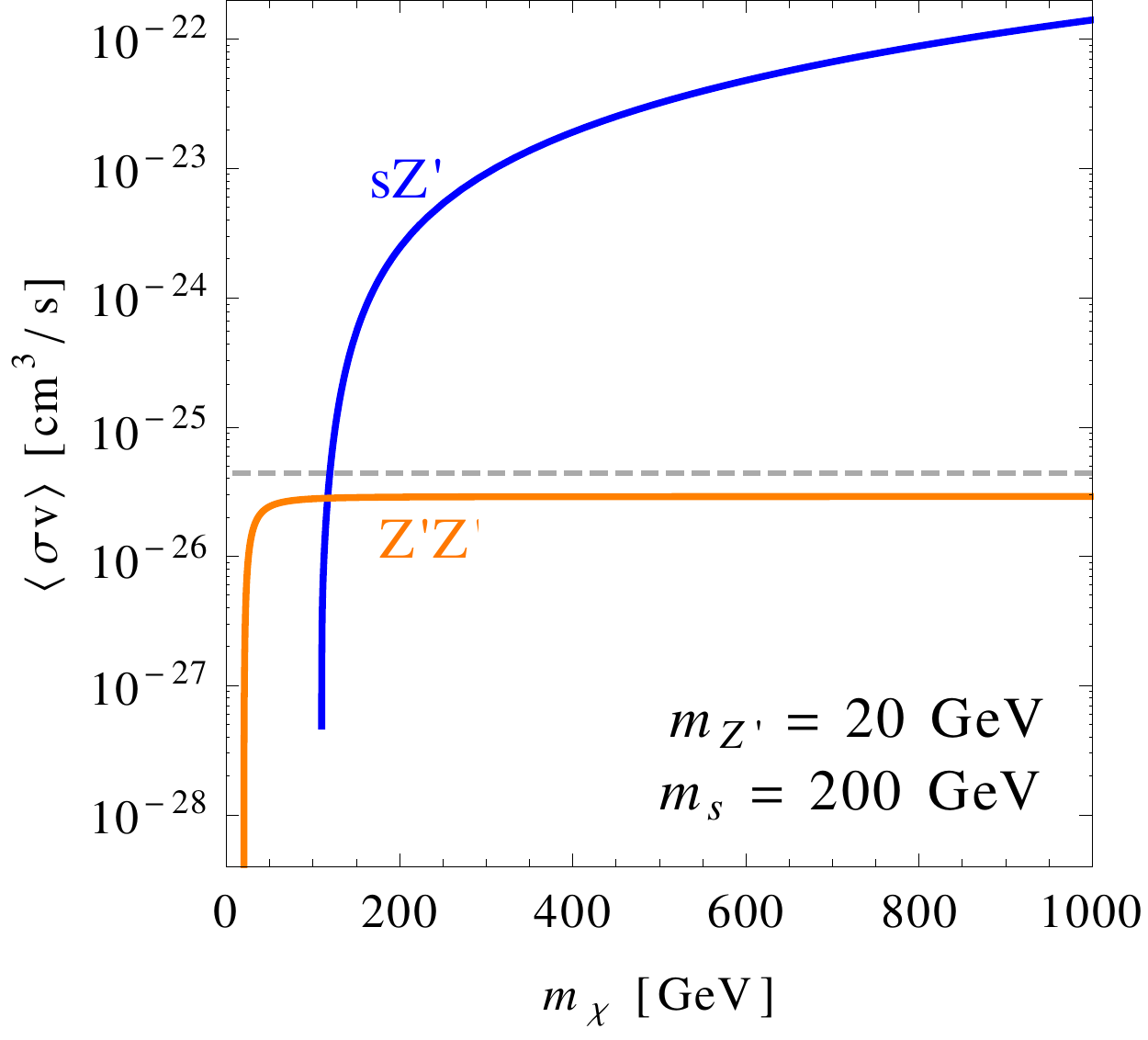}}
        \subfigure{
            \includegraphics[width=0.32\columnwidth]{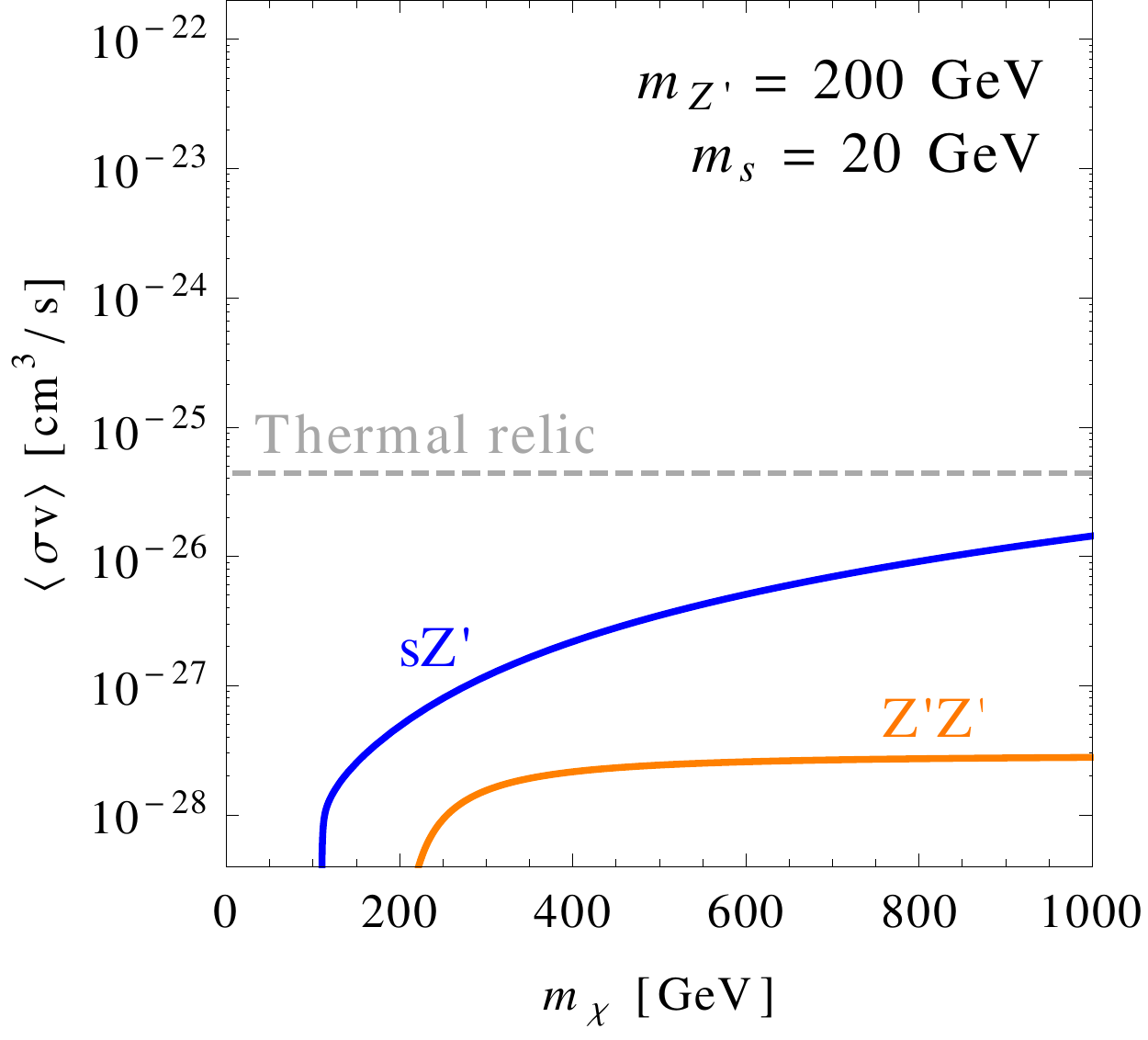}}
    \end{center}
    \caption{Relative cross sections for the two dominant $s$-wave
      annihilation processes in scenario II, $\chi\overline{\chi}\rightarrow s
      Z'$ (blue) and $\chi\overline{\chi}\rightarrow Z' Z'$ (orange), for
      example choices of the dark Higgs mass and the $Z'$ mass, as
      labeled.  This scenario requires $2Q_A = Q_S =1$, and we have
      chosen $Q_V=1/2$.  The gauge coupling has been set to
      $g_\chi=0.1$, but as all cross sections are directly
      proportional to $g_\chi^4$ they can easily be scaled by
      adjusting this parameter. For reference, the approximate thermal
      relic cross section is shown as the gray dashed line.}
   \label{fig:xsec_scenII}
\end{figure}

These annihilation cross sections are plotted in
Fig.~(\ref{fig:xsec_scenII}).  We see that the $Z'Z'$ cross section
becomes approximately independent of the DM mass when $m_\chi \gg
m_{Z'}$, while the $sZ'$ cross section rises with $m_\chi$.  (This is
to be contrasted with the behavior in cases III and IV, where all
cross sections decline as $m_\chi$ is increased.)  This is an
interesting consequence of having both vector and axial-vector
interactions present: For the $Z'Z'$ process, there is a $V-A$
interference which gives rise to longitudinal $Z'$ domination in the
$m_\chi^2 \gg m_{Z'}$ limit\footnote{If $Q_V$ were chosen to
  be zero, such that the $Z'$ coupling were pure axial, there would be
  no $V-A$ interference and the $s$-wave part of the $Z'Z'$ cross
  section would {\it not} be enhanced by longitudinal $Z'$ modes.
  This situation maps onto the Majorana DM case studied in
  Ref.~\cite{Bell:2016fqf}.}.  The $sZ'$ process is also dominated by
$Z'_L$ contributions in this limit.  This can be understood by
appealing to the Goldstone boson equivalence theorem which, in the
high energy limit, relates the amplitude for emission of a
longitudinally polarized gauge boson ($Z'_L$) with that for the
emission of the corresponding Goldstone boson (the pseudoscalar $a$).
For the $\chi\overline{\chi}\rightarrow sZ'$ process, in addition to the
transverse contributions we have $\chi\overline{\chi}\rightarrow sZ'_L$, which in
the high energy limit is equivalent to $\chi\overline{\chi}\rightarrow sa$. As
this scalar plus pseudoscalar final state is odd under parity, this is
an $s$-wave process. For the $\chi\overline{\chi}\rightarrow Z'Z'$ process, if
both $Z'_L$ are replaced by their Goldstones we would have
$\chi\overline{\chi}\rightarrow aa$, which is even under parity and thus
$p$-wave. However, a combination of longitudinal and transverse modes
are possible, $\chi\overline{\chi}\rightarrow Z'_LZ'_T$, which is equivalent to
the $s$-wave process $\chi\overline{\chi}\rightarrow aZ'_T$ and thus dominates at
high energy.  Notice that the $sZ'$ process, in addition to external
$Z'_L$ contributions, also receives contributions from the
longitudinal $Z'$ mode in the $s$-channel $Z'$ propagator.  This
contribution leads to four powers of $m_{Z'}$ in the denominator of
the $sZ'$ cross section.  In contrast, the $Z'Z'$ cross section
receives $Z'_L$ contributions only from a single final state $Z'$, and
so has only two powers of $m_{Z'}$ in the denominator.  The $Z'Z'$
process is thus is sub-dominant to the $sZ'$ process when both are
kinematically allowed\footnote{Note that because the Yukawa and gauge
  coupling constants are related via Eq.~(\ref{eq:coupling}), it is not
  possible to change the relative size of the annihilation to $Z'Z'$
  and $sZ'$ by adjusting these parameters.}.

\subsection{Relic Density}
\label{sec:relicdensII}

An important requirement for a DM model is to produce the
correct relic density.  Note, however, that a full DM model
is likely to have more dark sector fields than the simplified models
considered here, which may impact the relic density determination.
Nonetheless, we shall determine the relic density constraints for our
simplified models, to serve as a guide to the viable regions of
parameter space.

We use {\sc micrOMEGAs 3}~\cite{Belanger:2013oya} to calculate the
DM relic density, and compare with the recent determination
by the Planck collaboration~\cite{Ade:2013zuv},
\begin{align}
\Omega_\chi h^2 = 0.1196 \pm 0.0031.
\end{align}

\begin{figure}[h]
     \begin{center}
        \subfigure{
            \includegraphics[width=0.32\columnwidth]{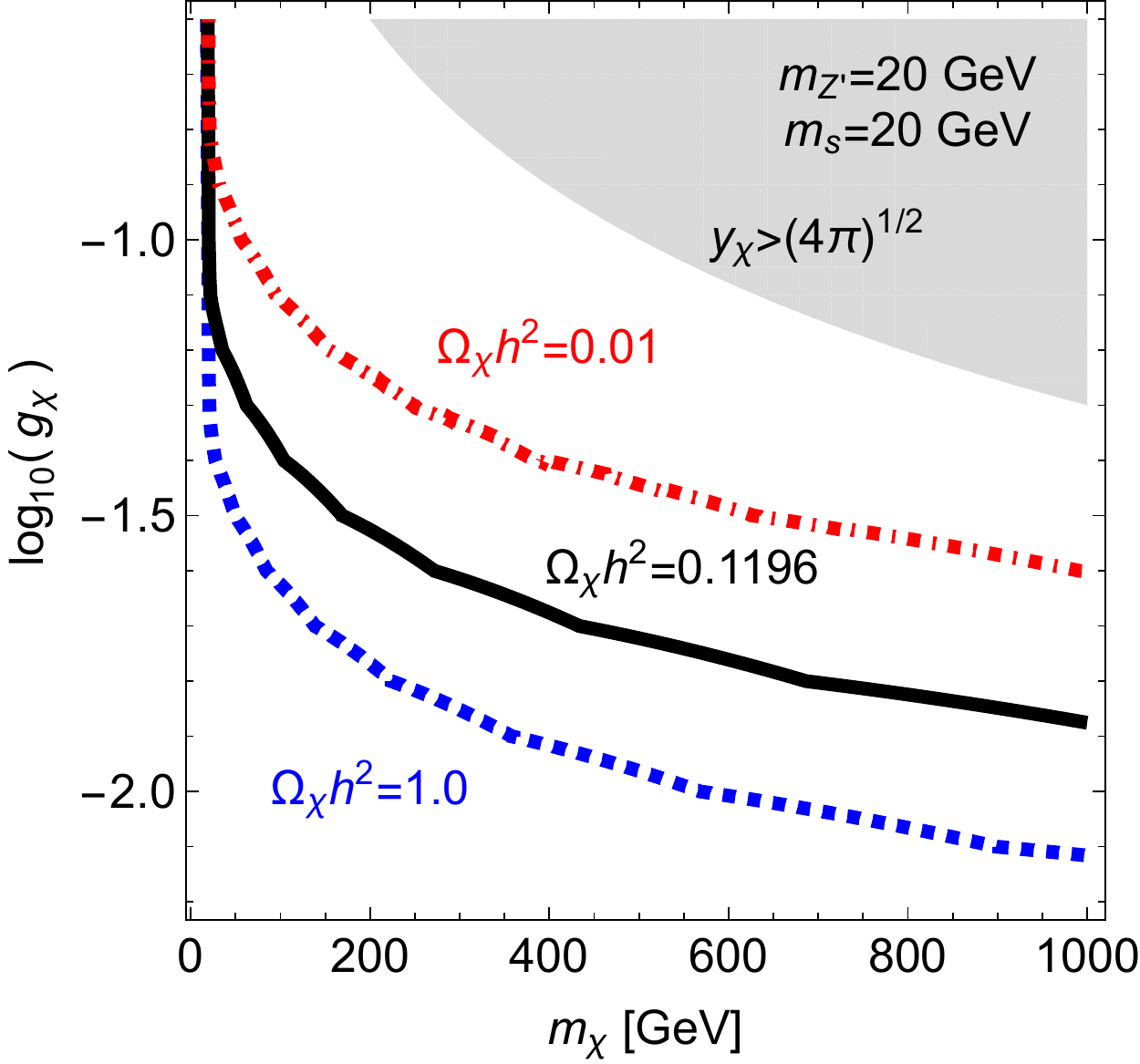}}
        \subfigure{
            \includegraphics[width=0.32\columnwidth]{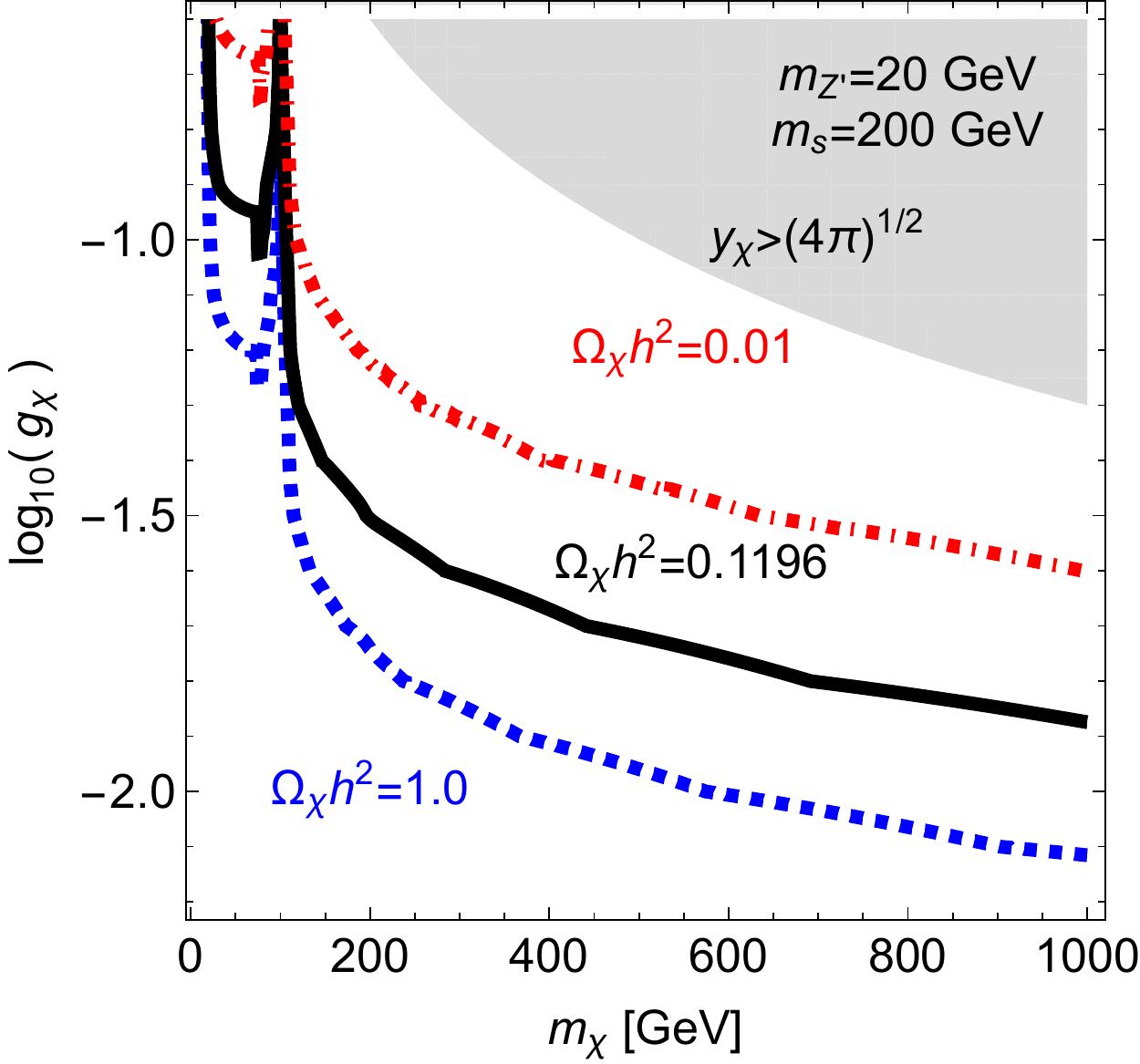}}
        \subfigure{
            \includegraphics[width=0.32\columnwidth]{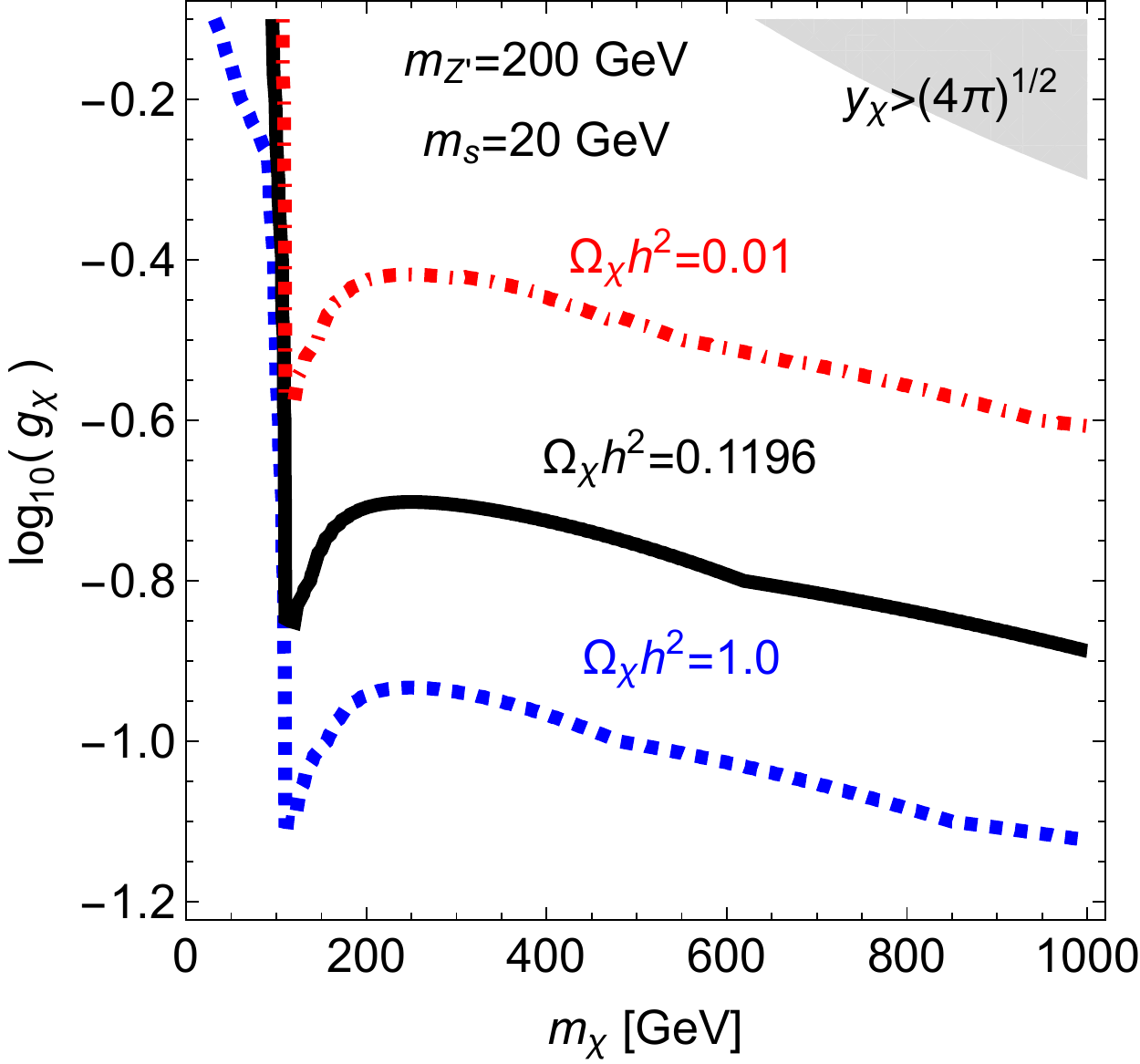}}
    \end{center}
\caption{Dark matter relic density contours in scenario II, as a
  function of $m_\chi$ and $g_\chi$, for various choices of the dark
  Higgs and $Z'$ mass, as labeled.  This scenario requires $2Q_A =
  Q_S =1$, and we have chosen $Q_V=1/2$. The red dot-dashed, the
  black solid and the blue dotted lines denote the contours for
  $\Omega_\chi h^2 = 0.01, \; 0.1196$ and $1.0$ respectively.  In the
  shaded region the Yukawa coupling $y_\chi$ is larger than
  $\sqrt{4\pi}$.}
\label{fig:relicII}
\end{figure}

For different $Z'$ and dark Higgs mass, we scan the parameter space
and find that the DM relic density can be saturated fairly
easily.  We show the relic contours in Fig.~(\ref{fig:relicII}) as a
function of $m_\chi$ and $g_\chi$ for fixed $m_{Z'}$ and $m_s$, and
$Q_V=1/2$.  In each panel, the observed relic $\Omega_\chi h^2 = 0.1196$
is depicted by a black solid line, while red dot-dashed and blue
dotted lines show contours for $\Omega_\chi h^2 = 0.01$ and $1.0$
respectively.  The central panel clearly shows the resonant
enhancement of the annihilation to $Z'Z'$ through the $s$-channel
scalar exchange, as a spike near $m_\chi\sim 100$ GeV.  Other features
of the relic contours in the middle and right panels are associated
with the $sZ'$ final state.  The parameter regions shown in
Fig.~(\ref{fig:relicII}) all satisfy $\lambda_s<\sqrt{4\pi}$.
 
\section{Scenario III: DM Mass from Dark Higgs Mechanism, $Z'$ Mass from Stueckelberg Mechanism}
\label{sec:scenIII}

\vspace{2mm}
\centerline{\textit{Interaction type required: Pure Vector}}
\vspace{4mm}

We now consider a scenario where the mass of the $\chi$ and $Z'$ arise
from different mechanisms.  Specifically, we assume the $\chi$ mass is
due to a Higgs mechanism, while the $Z'$ mass arises from the
Stueckelberg mechanism.  As a result only pure vector interactions
of the $\chi$ and $Z'$ are permitted.  Here the dark $U(1)_\chi$
remains unbroken, and instead the dark Higgs must break some other
symmetry under which the DM is charged. This scenario divorces the
$Z'$ physics from the dark Higgs physics\footnote{This situation has a
  SM analogue where $\chi$ is replaced by the electron and $Z'$ is
  replaced by the photon: the electrons have vector couplings to the
  photon of an unbroken $U(1)_{\rm QED}$; the electron mass comes from
  breaking the electroweak symmetry with the SM Higgs; the SM Higgs
  does not couple to the photon or contribute to the photon mass.  The
  annihilation $\chi\overline{\chi}\rightarrow sZ'$ is then the analogue of
  $e^+e^-\rightarrow h\gamma$.}.

\subsection{Model}
The most minimal Lagrangian for this setup is
\begin{eqnarray}
 \mathcal{L} = \mathcal{L}_{SM}
&+& i\, \overline{\chi}\left( \slashed{\partial} +ig_\chi Q_V \slashed{Z'} \right) \chi 
-\frac{y_\chi}{\sqrt{2}} \overline{\chi} {\chi} \phi  
\\
 &+& \frac{1}{2}\partial_\mu \phi \partial^\mu \phi 
-\frac{1}{2} \mu_s^2 \phi^2 
- \frac{1}{4}\lambda_s \phi^4
 - \frac{1}{2}\lambda_{hs} \phi^2 (H^\dagger H) 
-\frac{\sin\epsilon}{2} Z'^{\mu\nu}B_{\mu\nu},
\nonumber
\end{eqnarray}
with the real scalar $ \phi = w + s$, where $w$ is the vev of $\phi$
and $s$ is the dark Higgs.  The vectorlike charge $Q_V$ can be chosen
freely.  Again, the dark sector interacts with the visible sector in
two ways: via kinetic mixing or Higgs mass mixing.

As the dark Higgs is responsible only for generating fermion masses, a
real scalar is sufficient to accomplish this task. (The dark Higgs must
break the $U(1)_\chi$ in all other scenarios we consider, which
requires a complex scalar.)  If we introduce a complex scalar instead,
the extra degree of freedom will be a massless Goldstone boson and
will contribute to the radiation energy density of the universe.  If the
Goldstones had the same temperature as the SM neutrinos, they would
make a contribution equivalent to $N^\nu_{\rm eff}=4/7$, in marginal
agreement with current experimental observations. However, their
contribution to $N^\nu_{\rm eff}$ would be suppressed if they decoupled
early enough to not be heated by the annihilations of some SM
species~\cite{Weinberg:2013kea}.

\subsection{Cross Sections}

As shown in Tab.~(\ref{table:cases}), both the $Z'Z'$ and $sZ'$
processes receive contributions only from the $t/u$ channel diagrams,
as the absence of a $Z'$--$s$ interaction eliminates the $s$-channel
diagrams of scenario II.  The $s$-wave contributions to the
annihilation cross sections are given by
\begin{equation}
\langle\sigma v\rangle_{\chi\overline{\chi}\rightarrow Z' Z'} = \frac{g_{\chi}^4 Q_V^4 \left(1-\eta _{Z'}\right)^{3/2}}{4 \pi  m_{\chi }^2 \left(\eta _{Z'}-2\right)^2}
\end{equation}
and
\begin{equation}
\langle\sigma v\rangle_{\chi\overline{\chi}\rightarrow s Z'} = \frac{g_{\chi}^2 Q_V^2 y_{\chi }^2 \sqrt{\left(\eta _{s} - \eta _{Z'} - 4\right)^2 -16 \eta _{Z'}} 
\left(\left(\eta_{s}-\eta _{Z'}-4\right)^2+8\eta _{Z'}\right)}{64 \pi  m_{\chi }^2 \left(\eta _{s}+\eta _{Z'}-4\right)^2},
\end{equation}
where $\eta_{s,Z'}=m_{s,Z'}^2/m_\chi^2$.

\begin{figure}[h]
     \begin{center}
        \subfigure{
            \includegraphics[width=0.35\columnwidth]{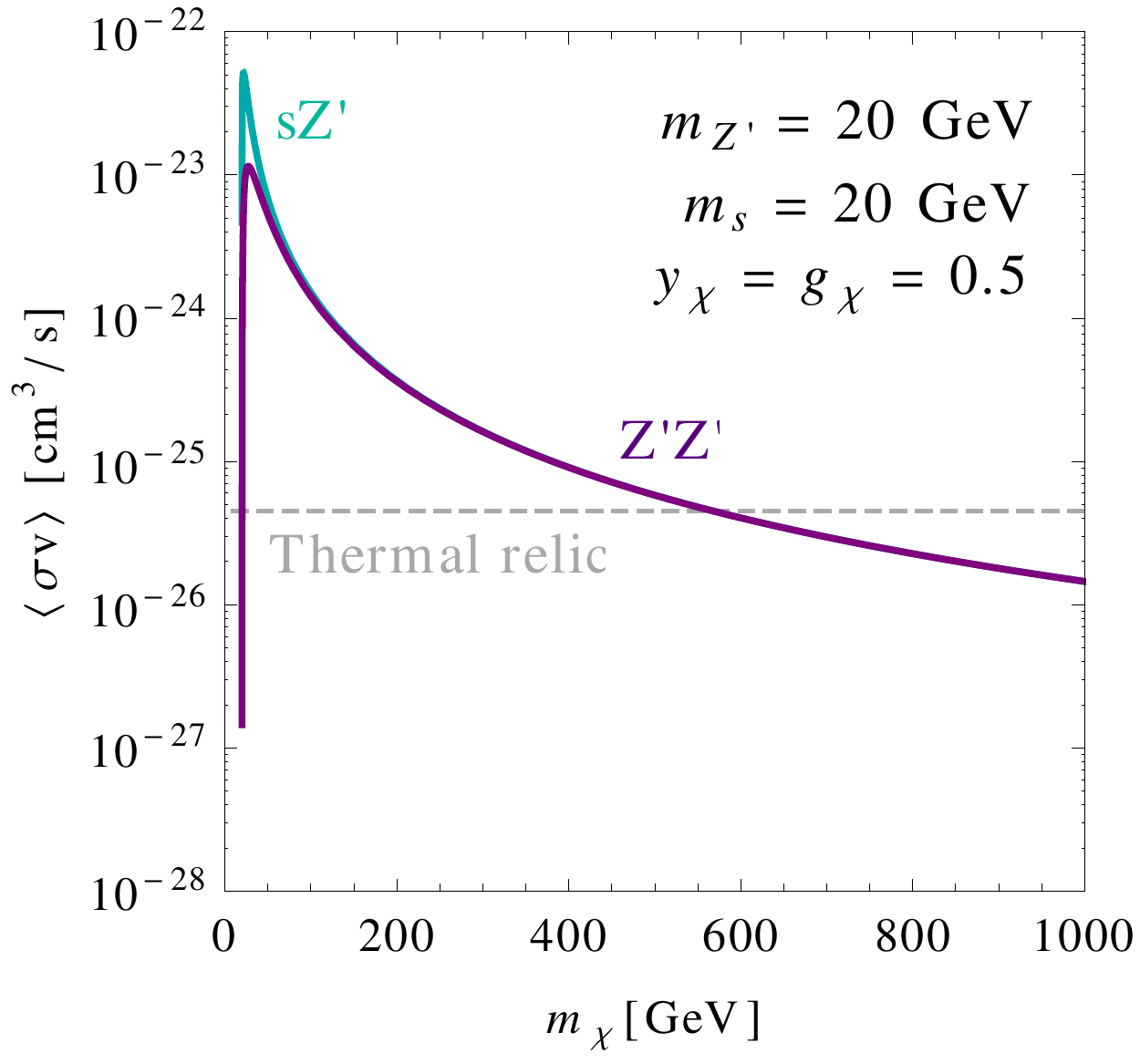}}
        \subfigure{
           \includegraphics[width=0.35\columnwidth]{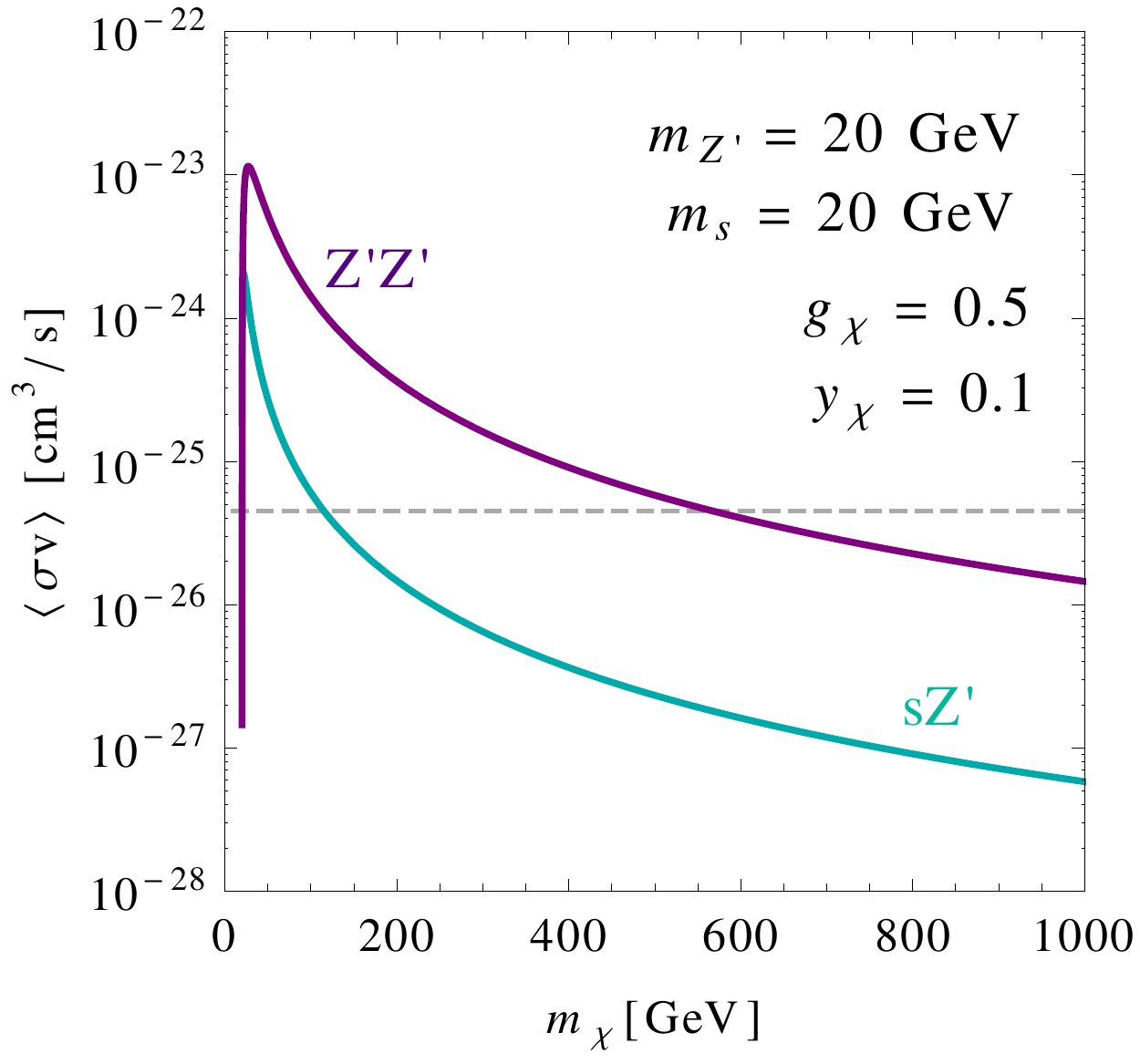}}
        \\ 
        \subfigure{
            \includegraphics[width=0.35\columnwidth]{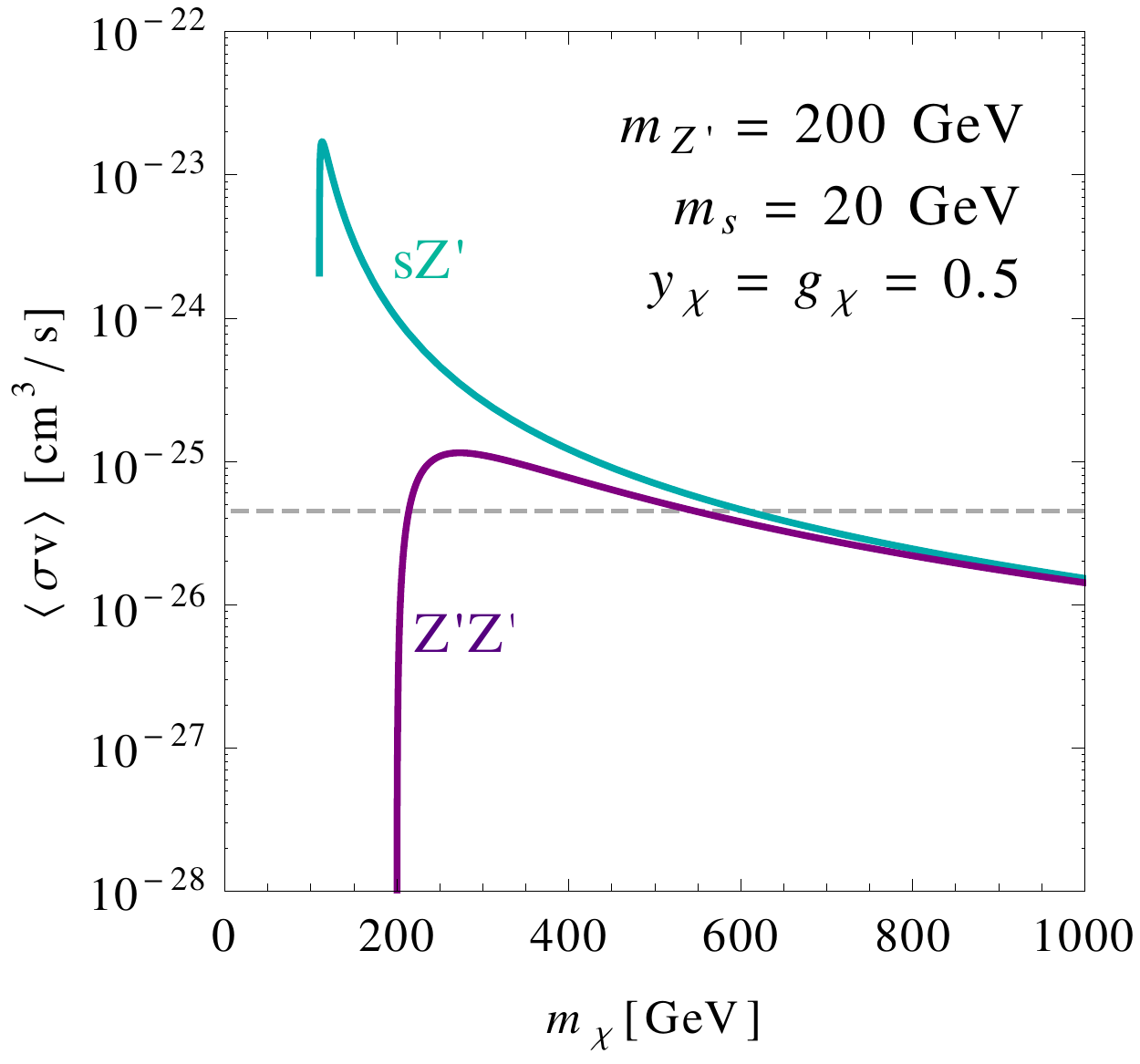}}
        \subfigure{
            \includegraphics[width=0.35\columnwidth]{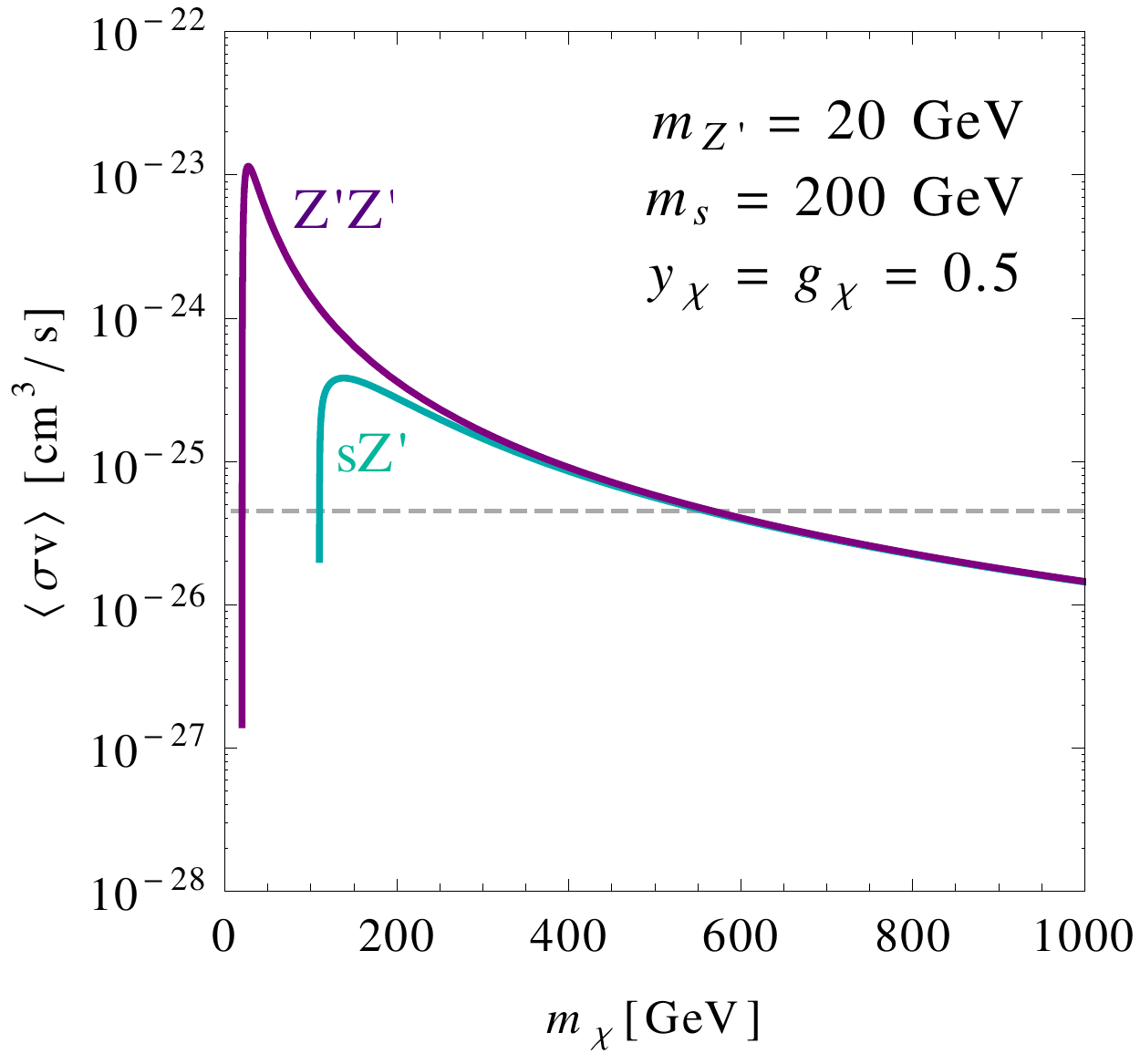}}
    \end{center}
    \caption{Cross section for the two dominant $s$-wave annihilation
      processes of scenario III, $\chi \overline{\chi}\rightarrow s Z'$ (blue)
      and $\chi \overline{\chi}\rightarrow Z' Z'$ (purple), for some example
      choices of the dark Higgs mass, the $Z'$ mass, and the dark
      gauge and Yukawa couplings, as labeled. Here $Q_V=1$. Notice, by comparing
      the top two panels, that either process can be chosen to
      dominate by varying the dark gauge coupling and the dark Yukawa
      coupling.  The approximate thermal relic cross section is shown
      as the gray dashed line.}
   \label{fig:xsec_scenIII}
\end{figure}

\begin{figure}[h]
     \begin{center}
        \subfigure{
            \includegraphics[width=0.29\columnwidth]{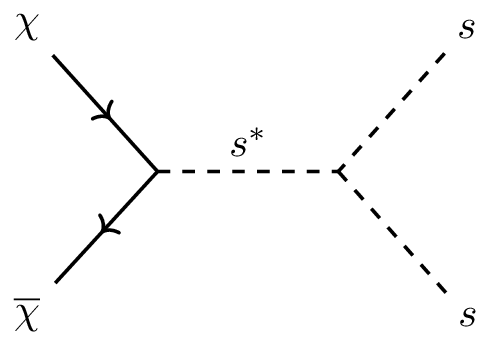}}\hspace{4mm}
        \subfigure{
            \includegraphics[width=0.32\columnwidth]{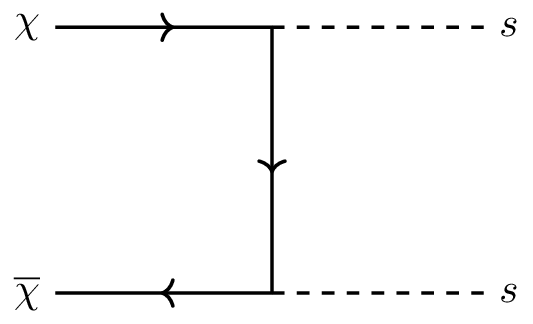}}
    \end{center}
    \caption{DM annihilation to two dark Higgs bosons, $\chi\overline\chi
      \rightarrow ss$. Despite being $p$-wave suppressed, these
      processes can make a non-negligible impact on the relic density
      at freeze-out, particularly if the gauge coupling is sufficiently
      small to suppress the $s$-wave processes.}
   \label{fig:ss_diagrams}
\end{figure}

The relative size and behavior of these cross sections can be seen in
Fig.~(\ref{fig:xsec_scenIII}).  Given that the $Z'$ obtains mass
from the Stueckelberg mechanism, there are no contributions to
the cross sections from longitudinal $Z'$ modes.  Therefore, all cross
sections decrease with increasing DM mass.  It is possible to dial the
strength of one annihilation process relative to the other by
adjusting the dark Yukawa coupling $y_\chi$ and the dark gauge
coupling $g_\chi$, which are independent parameters.  (This freedom
was not available in scenario II, where the couplings were related.)
This is shown in the top two panels of Fig.~(\ref{fig:xsec_scenIII}).
This also means that if $g_\chi \ll y_\chi$ then $p$-wave processes
such as $\chi\overline{\chi}\rightarrow ss$ (which scale as $y_\chi^4$) may have
an important effect on the relic density, as the otherwise dominant
$Z'Z'$ and $sZ'$ processes (which scale as $g_\chi^4$ and
$g_\chi^2y_\chi^2$ respectively) would be suppressed by the small
gauge coupling.  
However, it is difficult to make the annihilation to
$ss$ dominate in the universe today, where the $p$-wave modes are
suppressed by $v_\chi^2\approx10^{-6}$. To do so would require
$g_\chi^2 \sim 10^{-6} y_\chi^2 $ which, while possible, is a very
tuned scenario that we shall not consider.
The relevant diagrams for annihilation to $ss$ are shown in
Fig.~(\ref{fig:ss_diagrams}).

\subsection{Relic Density}

\begin{figure}[h]
     \begin{center}
        \subfigure{
            \includegraphics[width=0.32\columnwidth]{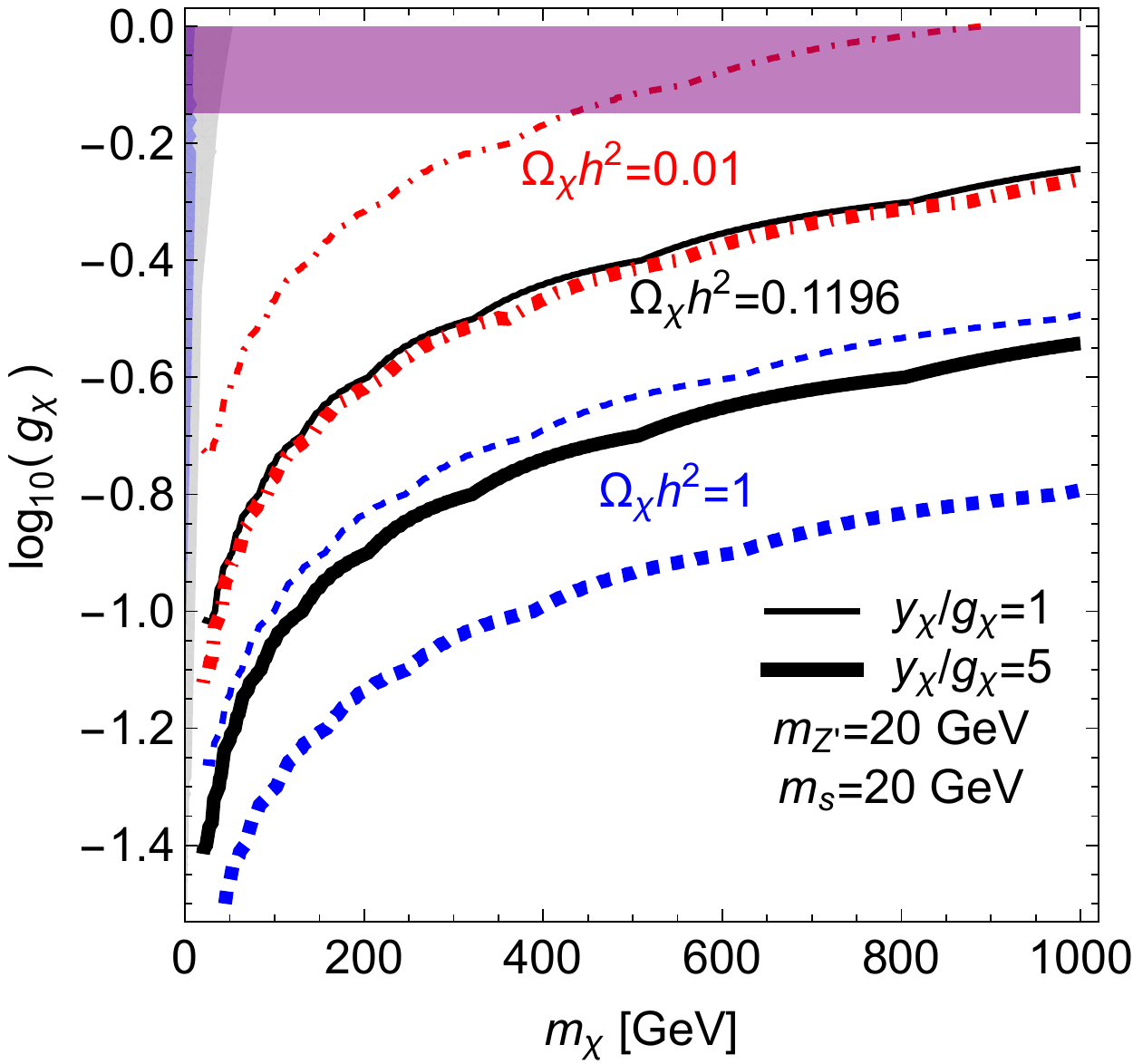}}
        \subfigure{
            \includegraphics[width=0.32\columnwidth]{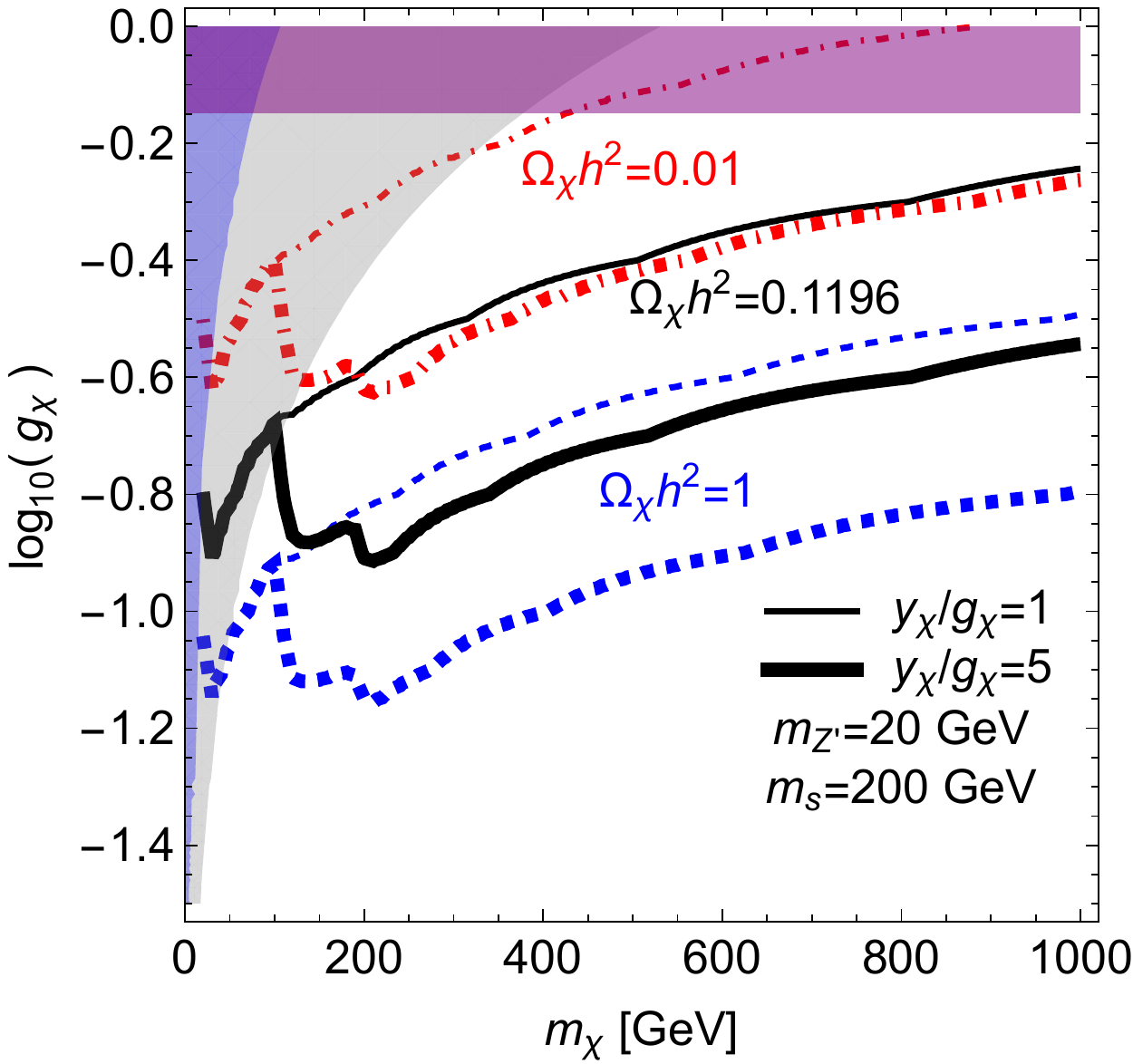}}
        \subfigure{
            \includegraphics[width=0.32\columnwidth]{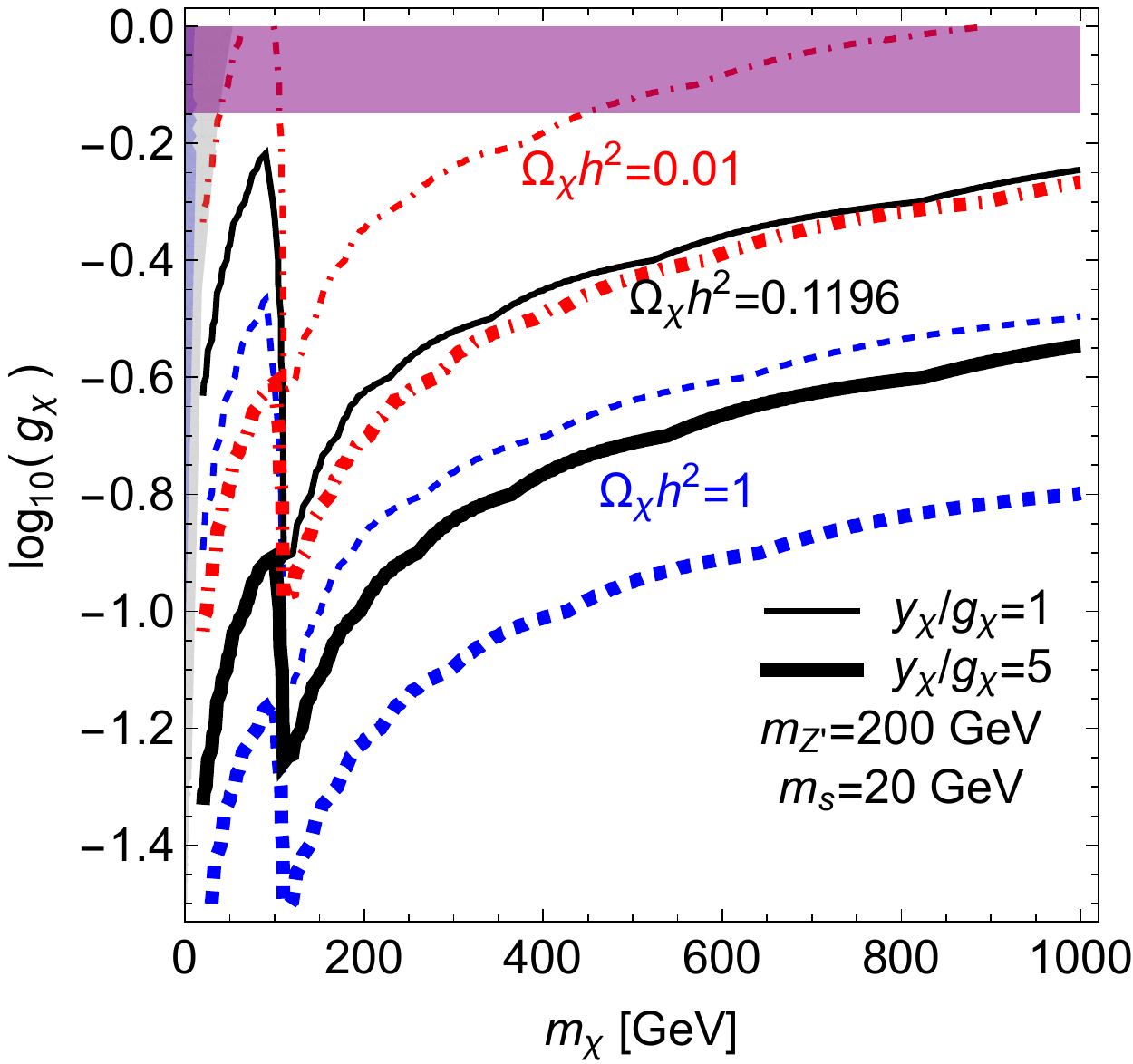}}
    \end{center}
\caption{Relic density contours for scenario III, as a function of
  $m_\chi$ and $g_\chi$, for various choices of the dark Higgs and
  $Z'$ mass and ratio of couplings constants, as labeled.  The thin
  (thick) red dot-dashed, solid black and dotted blue lines denote
  $\Omega h^2 = 0.01, 0.1196$ and $1$, respectively, for $y_\chi /
  g_\chi =1 \; (5)$.  In the light purple shaded regions, the Yukawa
  coupling $y_\chi$ is larger than $\sqrt{4\pi}$ for
  $y_\chi/g_\chi=5$; it is always perturbative in the parameter space
  shown when $y_\chi/g_\chi=1$.  The region shaded blue (gray)
  on the left side of each panel shows the perturbativity bound for
  $\lambda_{s}$ with $y_\chi/g_\chi=1 \;(5)$ .  }
\label{fig:relic2}
\end{figure}

In Fig.~(\ref{fig:relic2}), we show the relic density contours as a
function of DM mass $m_\chi$ and the dark gauge coupling
$g_\chi$ for various values of the $Z'$ mass, dark Higgs mass, $Q_V=1$ and fixed
ratios of $y_\chi/g_\chi$.  The color codes for the contours are the
same as in the previous scenario.  The different choices of
$y_\chi/g_\chi$ are embodied in the thickness of the lines: thinner
for $y_\chi/g_\chi=1$ and thicker for $y_\chi/g_\chi=5$.  Obviously
for the same $g_\chi$, a larger $y_\chi/g_\chi$ ratio results in a
larger cross section for $\chi\overline{\chi}\to sZ'$ and thus a smaller relic
density; a smaller $g_\chi$ is thus needed to obtain the same relic
density, resulting in an overall downward shift of the contours.  In
this scenario, the quartic coupling $\lambda_s$ can be expressed as
$\lambda_s\simeq y_\chi^2 m_s^2/(4 m_\chi^2)$.  The perturbativity
bounds for $\lambda_s$ are shown by the shaded gray regions, while the
parameter space where $y_\chi > \sqrt{4\pi}$ ($y_\chi/g_\chi =5$ only)
is shown as the light purple region.

The dips in Fig.~(\ref{fig:relic2}) correspond to kinematic opening of
various channels when $2m_\chi > m_1 + m_2$ with $m_{1,2}$ being
the masses of the annihilation product.  For $2m_\chi< m_1+m_2$,
the annihilation of DM through these channels will be
exponentially suppressed.  So in the left panel, where $m_{Z'}=m_s=20$
GeV, all three channels, $sZ'$, $Z'Z'$ and $ss$, are open at the same
time and no dips occur beyond this mass.  For the middle and right
panels of Fig.~(\ref{fig:relic2}), there is a dip at $m_\chi \simeq 110$
GeV corresponding to the $sZ'$ channel.  In the middle panel, the $ss$
channel begins to contribute around $m_\chi\simeq 200$ GeV; the effect is
more pronounced for larger $y_\chi$, leading to a prominent dip for
$y_\chi/g_\chi=5$ but not for $y_\chi/g_\chi=1$.  In the right panel,
however, there is no dip around 200 GeV for the $Z'Z'$ channel, since
the $Z'Z'$ cross section is always subdominant to the $sZ'$ cross section
for the couplings chosen.

\section{Scenario IV: Bare DM Mass, $Z'$ Mass from Dark Higgs Mechanism}
\label{sec:scenIV}

\vspace{2mm}
\centerline{\textit{Interaction type required: Pure Vector}}
\vspace{4mm}

An alternative scenario in which the mass of the DM and $Z'$ arise
from different mechanisms, is to have a bare mass for the $\chi$ and
use a dark Higgs mechanism to provide mass for the $Z'$.  In this
scenario, again, only pure vector interactions of the $\chi$ and $Z'$
are permitted.

\subsection{Model}

In this scenario, the most minimal gauge invariant Lagrangian is 
\begin{eqnarray}
 \mathcal{L}&= & \mathcal{L}_{SM} +  
i\, \overline{\chi}\left( \slashed{\partial} +ig_\chi Q_V \slashed{Z'}\right) \chi 
-\frac{\sin\epsilon}{2} Z'^{\mu\nu}B_{\mu\nu}-m_\chi\overline{\chi}\chi\\
 &+& \left[ (\partial^\mu + ig_\chi Q_S Z'^\mu ) S \right]^\dagger \left[ (\partial_\mu + ig_\chi Q_S Z'_\mu ) S \right] 
- \mu_s^2 S^\dagger S - \lambda_s (S^\dagger S)^2
 - \lambda_{hs}(S^\dagger S)(H^\dagger H). \nonumber
\end{eqnarray}
The vectorlike charge $Q_V$ and dark Higgs charge $Q_S$ under the dark
$U(1)_\chi$ can be chosen freely. Again the dark sector interacts with
the visible sector in two ways: via kinetic mixing or Higgs mass
mixing.

\subsection{Cross Sections}

\begin{figure}[h]
     \begin{center}
        \subfigure{
            \includegraphics[width=0.35\columnwidth]{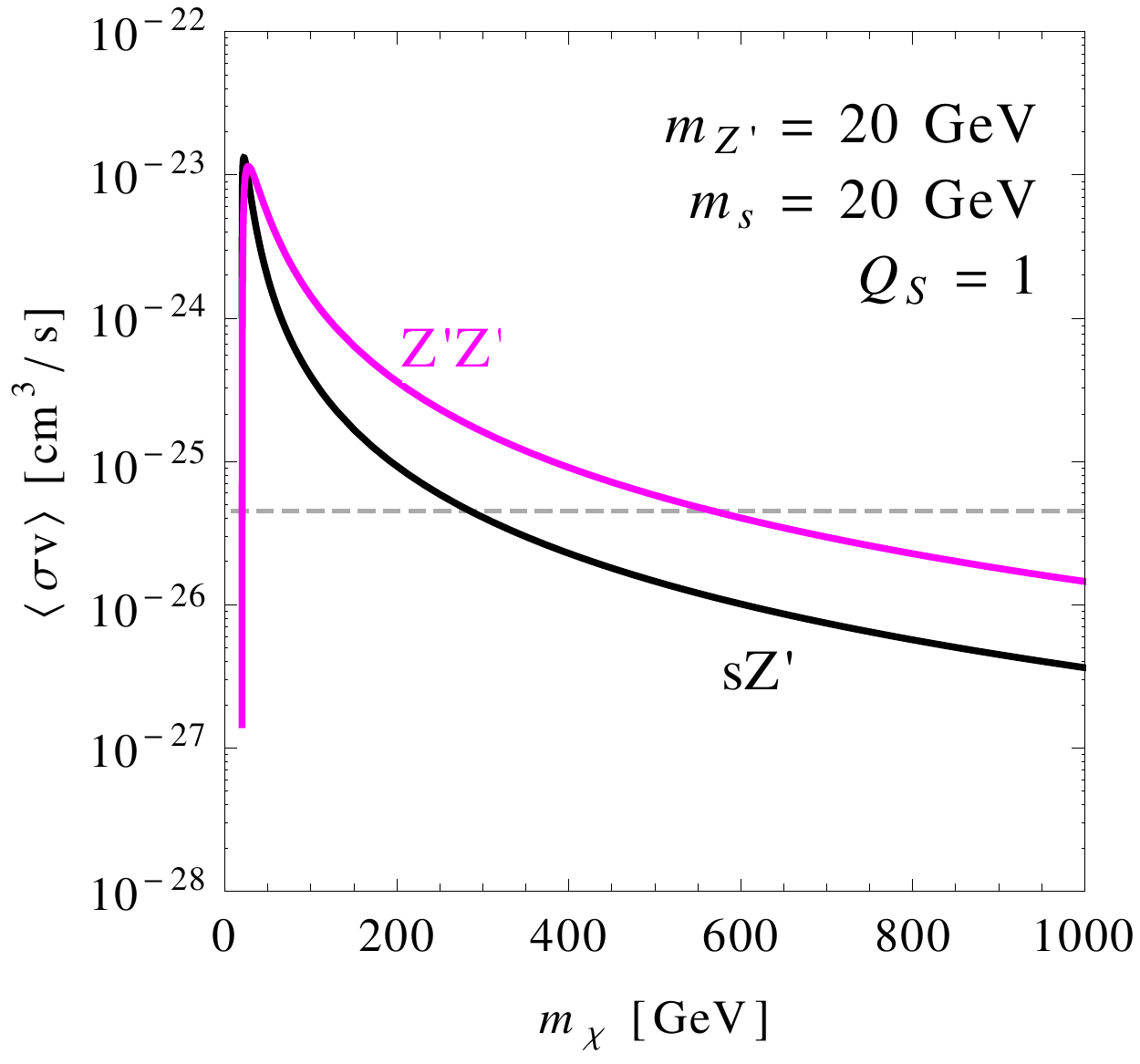}}
        \subfigure{
           \includegraphics[width=0.35\columnwidth]{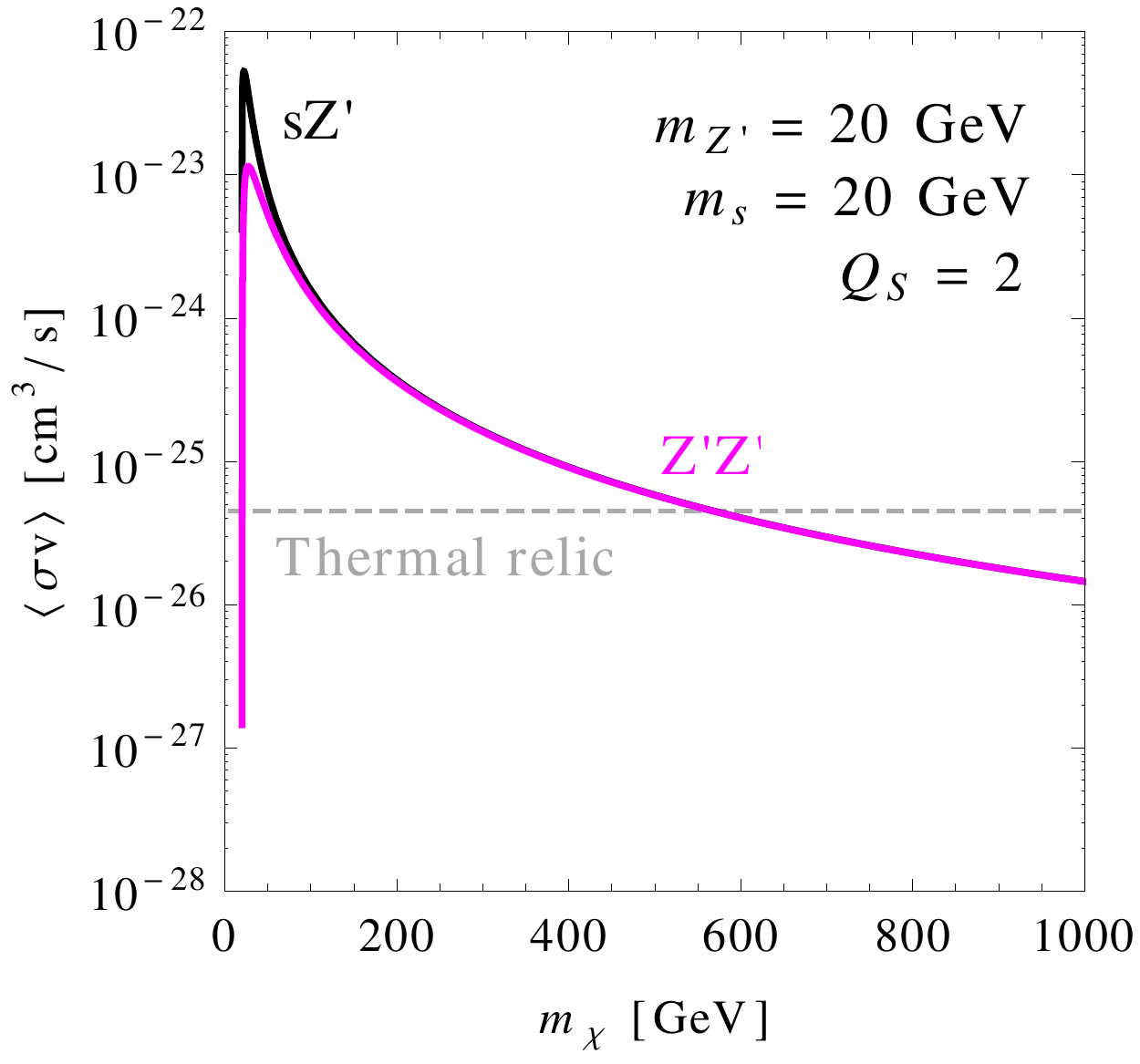}}
        \\ 
        \subfigure{
            \includegraphics[width=0.35\columnwidth]{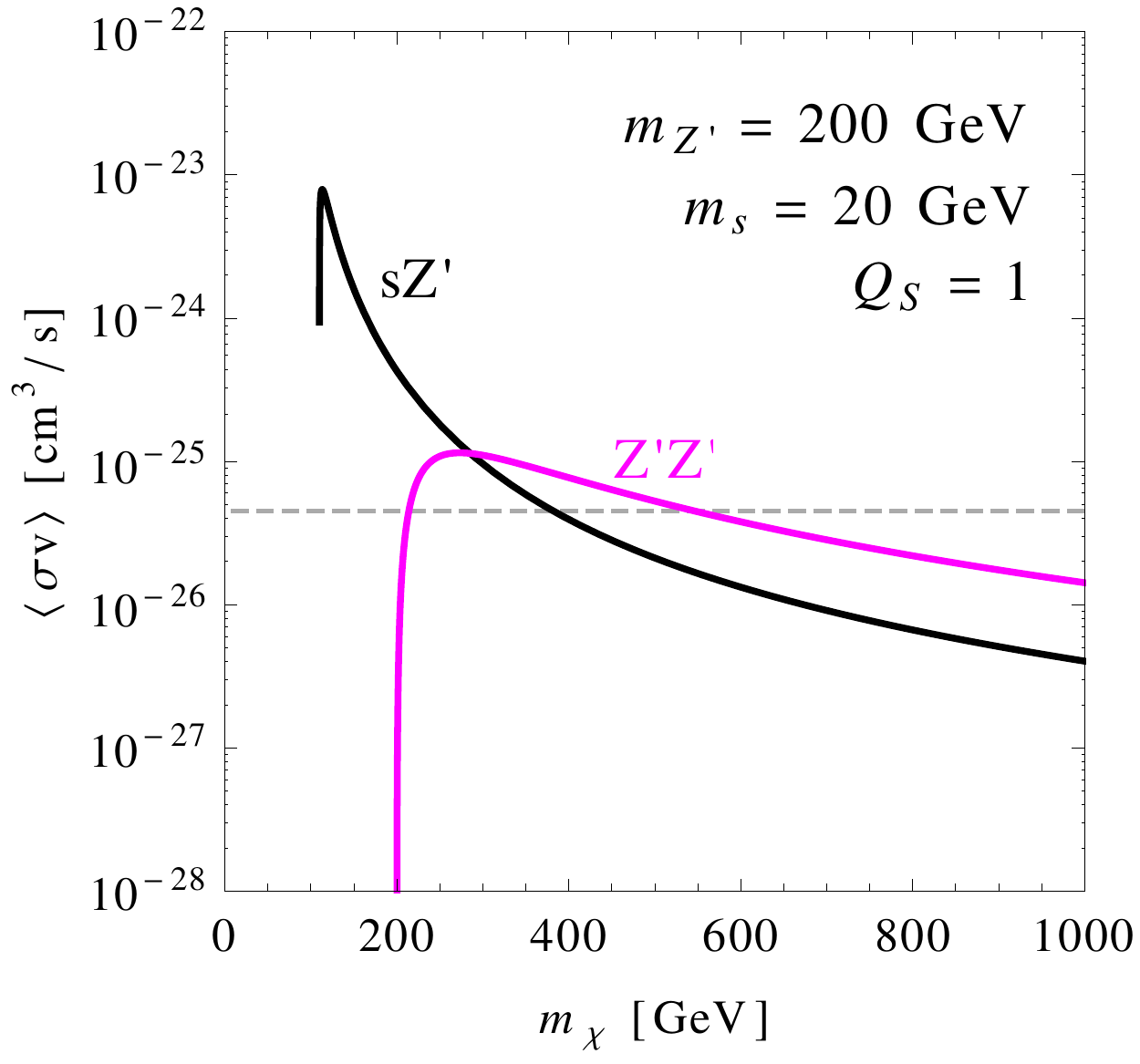}}
        \subfigure{
            \includegraphics[width=0.35\columnwidth]{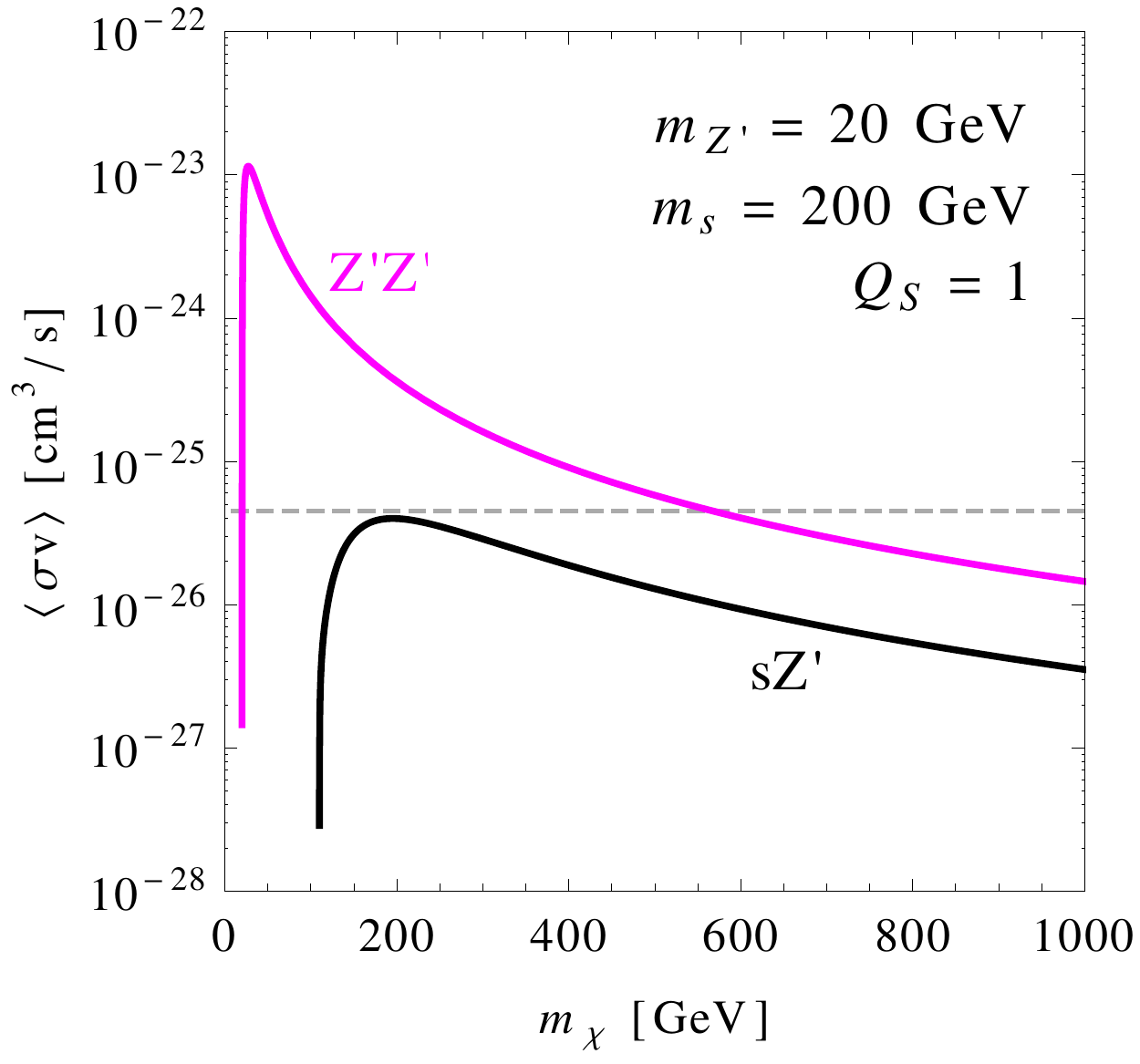}}
    \end{center}
    \caption{Relative cross section for the two dominant $s$-wave
      annihilation processes in scenario IV, $\chi\overline{\chi}\rightarrow s
      Z'$ (black) and $\chi\overline{\chi}\rightarrow Z' Z'$ (magenta), for
      some example parameter choices for the dark Higgs mass and $Z'$
      mass, as labeled. Here $Q_V=1$. Example dark charges $Q_S=$1,2 are shown, which demonstrate
  how either process can be made to dominate, or both can be made
  comparable, if kinematically allowed. For all plots the gauge
  coupling is set to $g_\chi=0.5$. As all cross sections are directly proportional to $g_\chi^4$ they can easily be scaled by adjusting this parameter.  The approximate thermal
relic cross section is shown as the gray dashed line.}
   \label{fig:xsec_scenIV}
\end{figure}

As shown in Tab.~(\ref{table:cases}), the annihilation to $sZ'$ proceeds
only via the $s$-channel diagram, as the DM does not interact directly
with the dark Higgs.  The annihilation to $Z'Z'$ proceeds via the
$t/u$ and channel diagrams.  The $s$-wave contributions to these
annihilation cross sections are given by
\begin{equation}
\langle\sigma v\rangle_{\chi \overline{\chi}\rightarrow Z' Z'} = \frac{g_\chi^4 Q_V^4 \left(1-\eta _{Z'}\right)^{3/2}}{4 \pi  m_{\chi }^2 \left(\eta _{Z'}-2\right)^2}
\end{equation}
and
\begin{equation}
\langle\sigma v\rangle_{\chi \overline{\chi}\rightarrow s Z'} =\frac{g_\chi^4 Q_V^2 Q_S^2 
\sqrt{ \left(\eta _s - \eta _{Z'} -4\right)^2 - 16 \eta_{Z'}} 
\left( \left(\eta _s - \eta _{Z'} - 4\right)^2+32\eta_{Z'}\right)}{256 \pi  m_{\chi }^2 \left(\eta _{Z'}-4\right)^2},
\end{equation}
where $\eta_{s,Z'}=m_{s,Z'}^2/m_\chi^2$. The behavior of these cross
sections is depicted in Fig. (\ref{fig:xsec_scenIV}).  We see that the
shapes of the $sZ'$ and $Z'Z'$ cross sections are similar, as both
fall off with DM mass as $1/m_\chi^2$.  There is no production of
longitudinal $Z'_L$ modes in the high energy limit, which is
consistent with the fact that the DM does not interact with Goldstone
modes, given the absence of a DM-Higgs coupling.  Because
$Q_{V}$ and $Q_S$ are independent, the relative size of the
$Z'Z'$ and $sZ'$ processes can again be scaled relative to each other
by appropriate choices of these charges.

\subsection{Relic Density}

\begin{figure}
     \begin{center}
        \subfigure{
            \includegraphics[width=0.32\columnwidth]{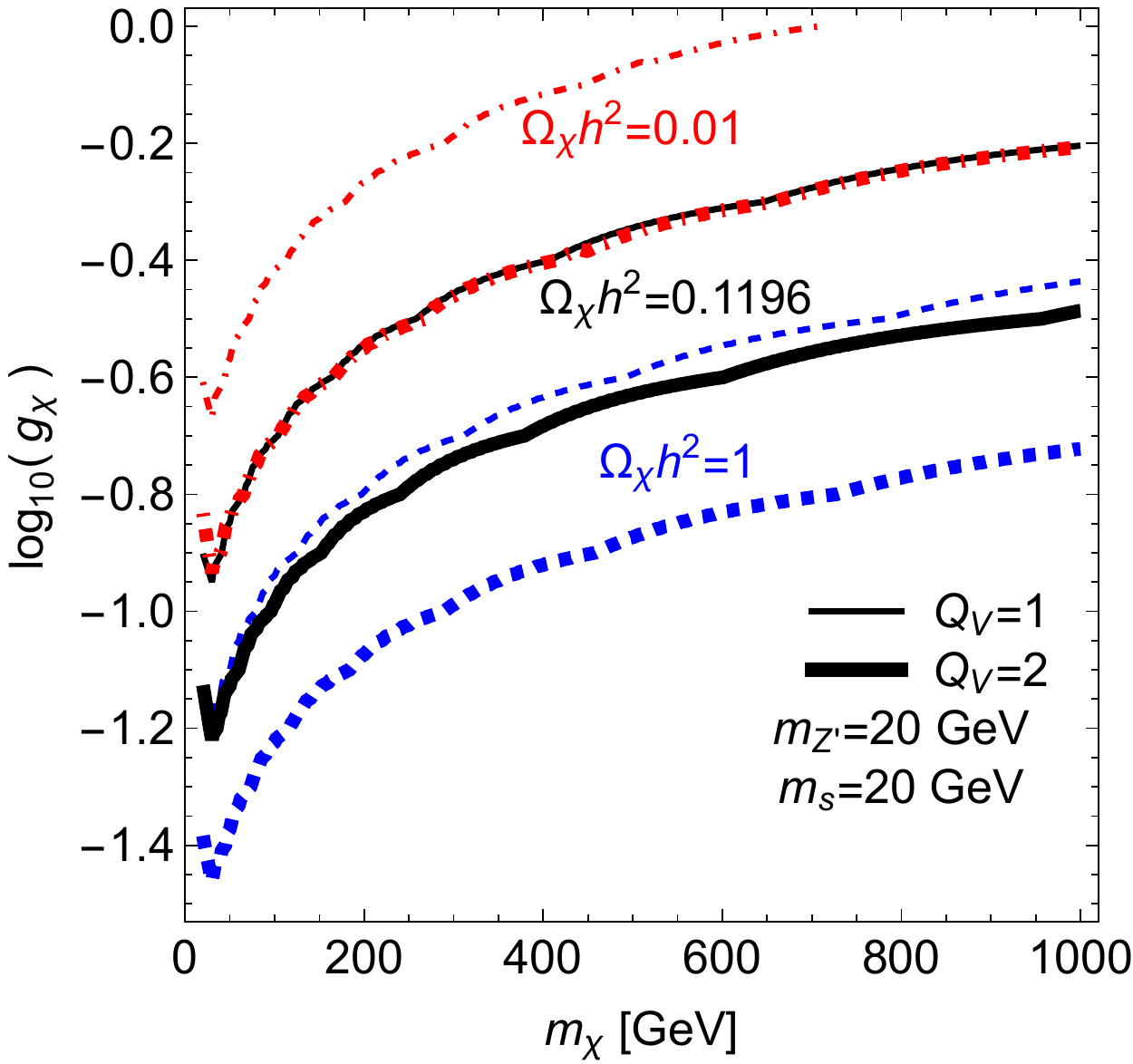}}
        \subfigure{
            \includegraphics[width=0.32\columnwidth]{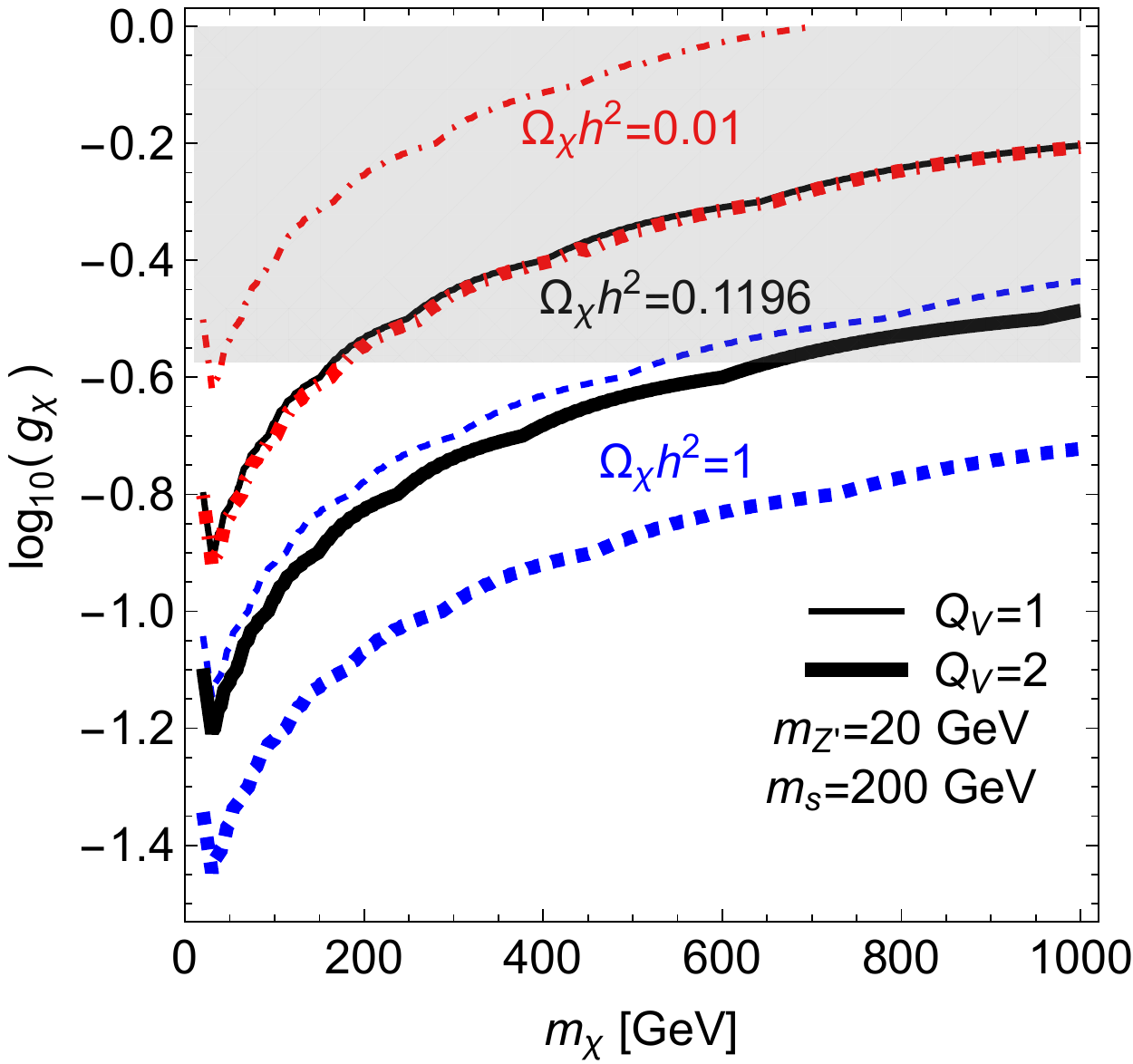}}
        \subfigure{
            \includegraphics[width=0.32\columnwidth]{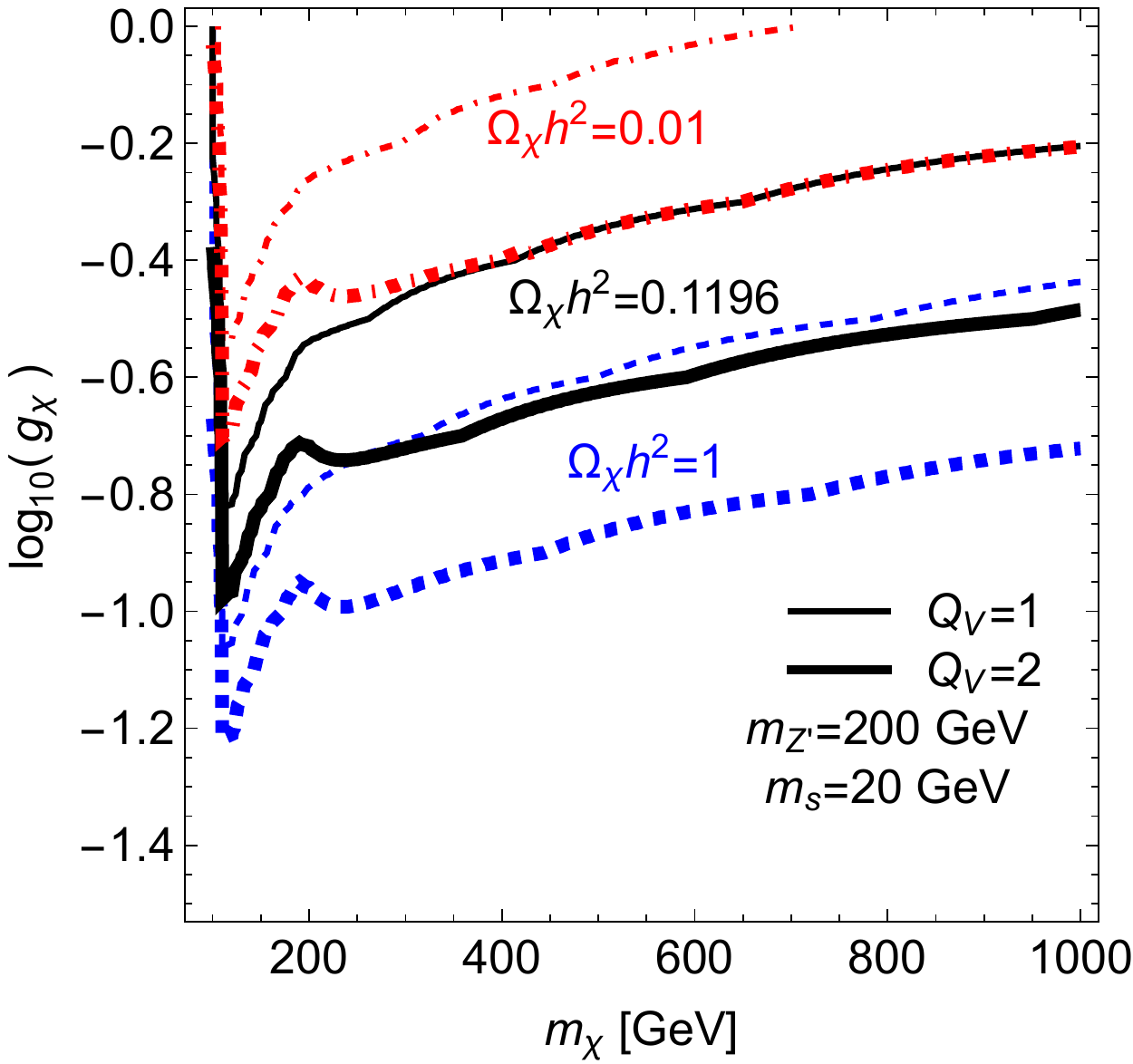}}
    \end{center}
\caption{Dark matter relic density contours for scenario IV as a
  function of $m_\chi$ and $g_\chi$, for various choices of $m_s$,
  $m_{Z'}$ and $Q_{V}$, as labeled.  The thin (thick) red dot-dashed,
  solid black and dotted blue lines denote $\Omega h^2 = 0.01, 0.1196$
  and $1$, respectively, for $Q_V =1 \; (2)$.   We have taken
    $Q_S=1$. For the chosen values of $Q_{V,S}$, the $sZ'$
  contribution is subdominant to that from $Z'Z'$ when both are
  kinematically allowed, which accounts for the features of the
  curves.  The light gray shaded region at the top of the middle panel
  shows the parameter space excluded by perturbativity bound for
  $\lambda_{s}$.}
\label{fig:relic3}
\end{figure}

We plot the relic density contours for this scenario in
Fig.~(\ref{fig:relic3}).  As the dark Higgs is responsible for the $Z'$
mass, the quartic coupling may be expressed as $\lambda_s\simeq
g_\chi^2 m_s^2/(2 m_{Z'}^2)$.  Parameters excluded by the
perturbativity bound on $\lambda_s$ are shaded gray; this bound is
relevant only for the middle panel of Fig.~(\ref{fig:relic3}), where the
ratio of $m_s/m_{Z'}$ is larger.  Because there is no direct coupling
of the scalar to the DM, there is no annihilation to $ss$.  As a
result, the features of the relic density contours are generally
simpler than in the previous scenario.  For the chosen values of
$Q_{V,S}$, the annihilation to $sZ'$ is subdominant to the $Z'Z'$
process when both are kinematically allowed.  This leads to a dip in
the contours of the right panel at $m_\chi \simeq m_{Z'}$, where the
$Z'Z'$ modes becomes allowed, but not in the left and center panels
where the $Z'Z'$ mode always plays the dominant role.

\section{Indirect Detection Phenomenology}
\label{sec:indirectdet}

We now determine indirect detection constraints on the dominant
annihilation modes for the scenarios discussed, $\chi \overline\chi
\rightarrow Z'Z'$ and $\chi \overline\chi \rightarrow sZ'$.  The $Z'$
and $s$ produced in these annihilations decay to SM particles, and
subsequent hadronization/decay of these SM states leads to gamma-ray
and other fluxes that we may compare with observational limits.

We generate our gamma-ray spectra as per the method outlined in
Ref.~\cite{Bell:2016fqf}, where a more detailed description can be
found.  The kinetic mixing of the $Z'$ with the SM hypercharge boson
permits the decay $Z'\rightarrow f\overline{f}$, with a partial width
given by
  \begin{equation}
   \Gamma(Z'\rightarrow f\bar{f})=\frac{m_{Z'}N_c}{12\pi}\sqrt{1-\frac{4m_f^2}{m_{Z'}^2}}\left[g_{f,V}^2\left(1+\frac{2m_f^2}{m_{Z'}^2}\right)+g_{f,A}^2\left(1-\frac{4m_f^2}{m_{Z'}^2}\right)\right],
   \label{eq:zppartial}
 \end{equation}
where $N_c$ is a color factor, relevant for hadronic decays.  The
$g_{f,V}$ (vector) and $g_{f,A}$ (axial-vector) structure of the
$Z'$--$f$ couplings are inherited from the kinetic
mixing~\cite{Agashe:2014kda}. The total decay width for the $Z'$ is
then simply given by the sum over all the final state fermions,
$\Gamma_Z'=\sum_f \Gamma(Z'\rightarrow f\bar{f})$.  The dark Higgs
decays to the SM due to mass mixing with the SM Higgs, and so it
decays preferentially to heavier particles. The dark Higgs is also
permitted to decay to pairs of $Z'$. In order to take into account
loop decays and higher order corrections, we calculate the dark Higgs
decay widths numerically with the {\sc Fortran} package {\sc
  HDecay}~\cite{Djouadi:1997yw}.

\begin{figure}[t]
\centering
\subfigure{\includegraphics[width=0.42\columnwidth]{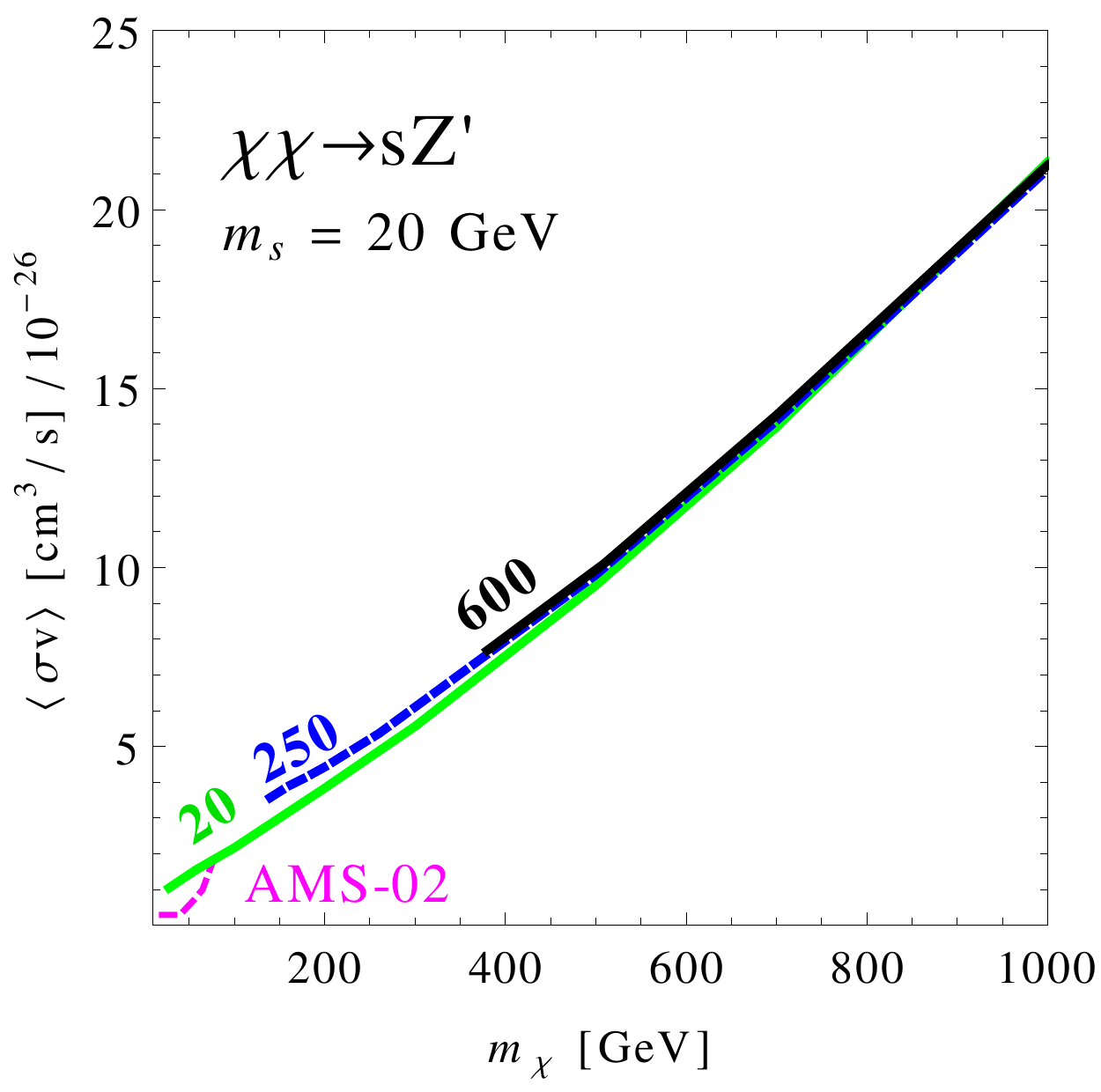}}
\subfigure{\includegraphics[width=0.42\columnwidth]{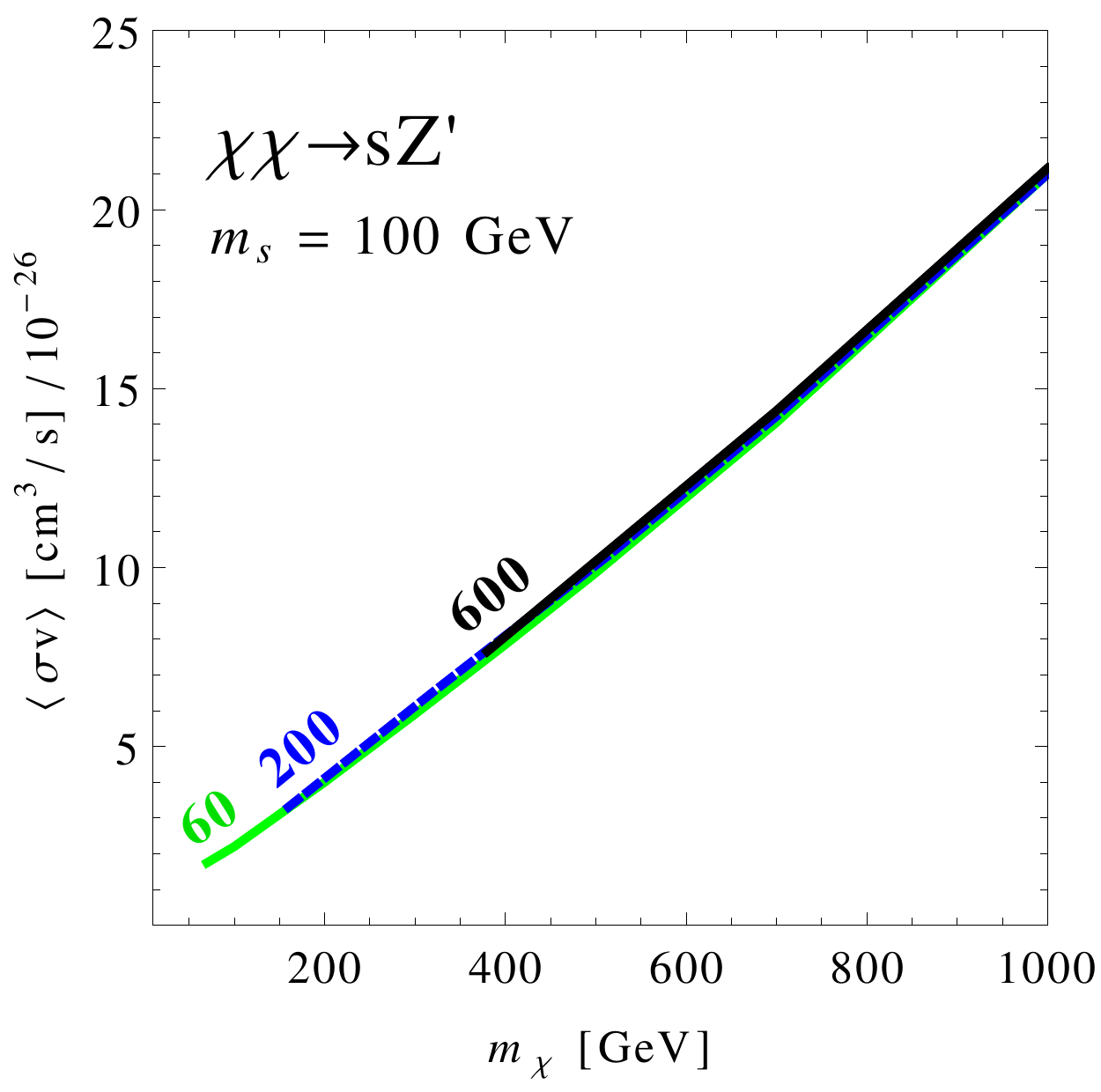}}\\
\subfigure{\includegraphics[width=0.42\columnwidth]{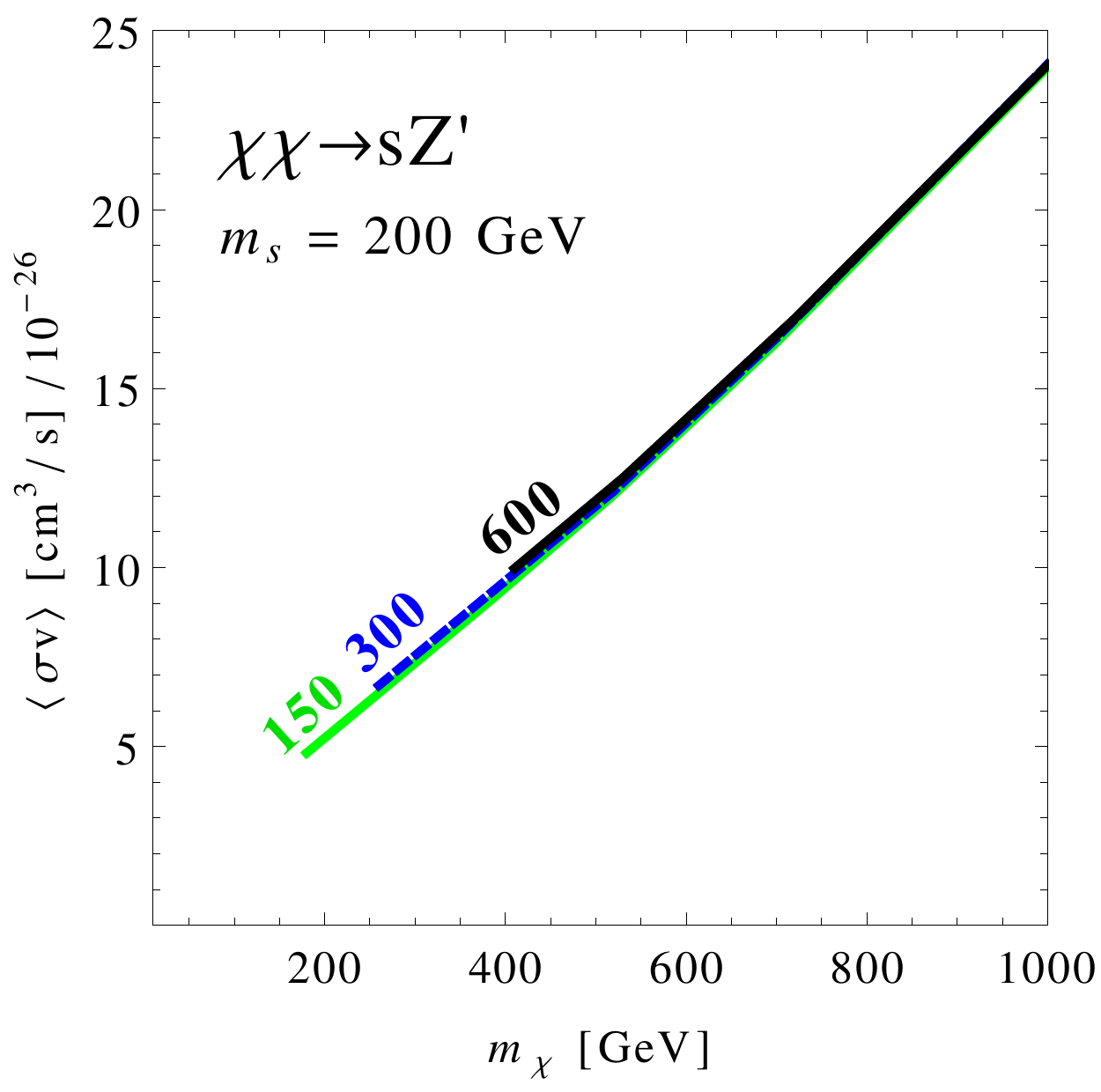}}
\subfigure{\includegraphics[width=0.42\columnwidth]{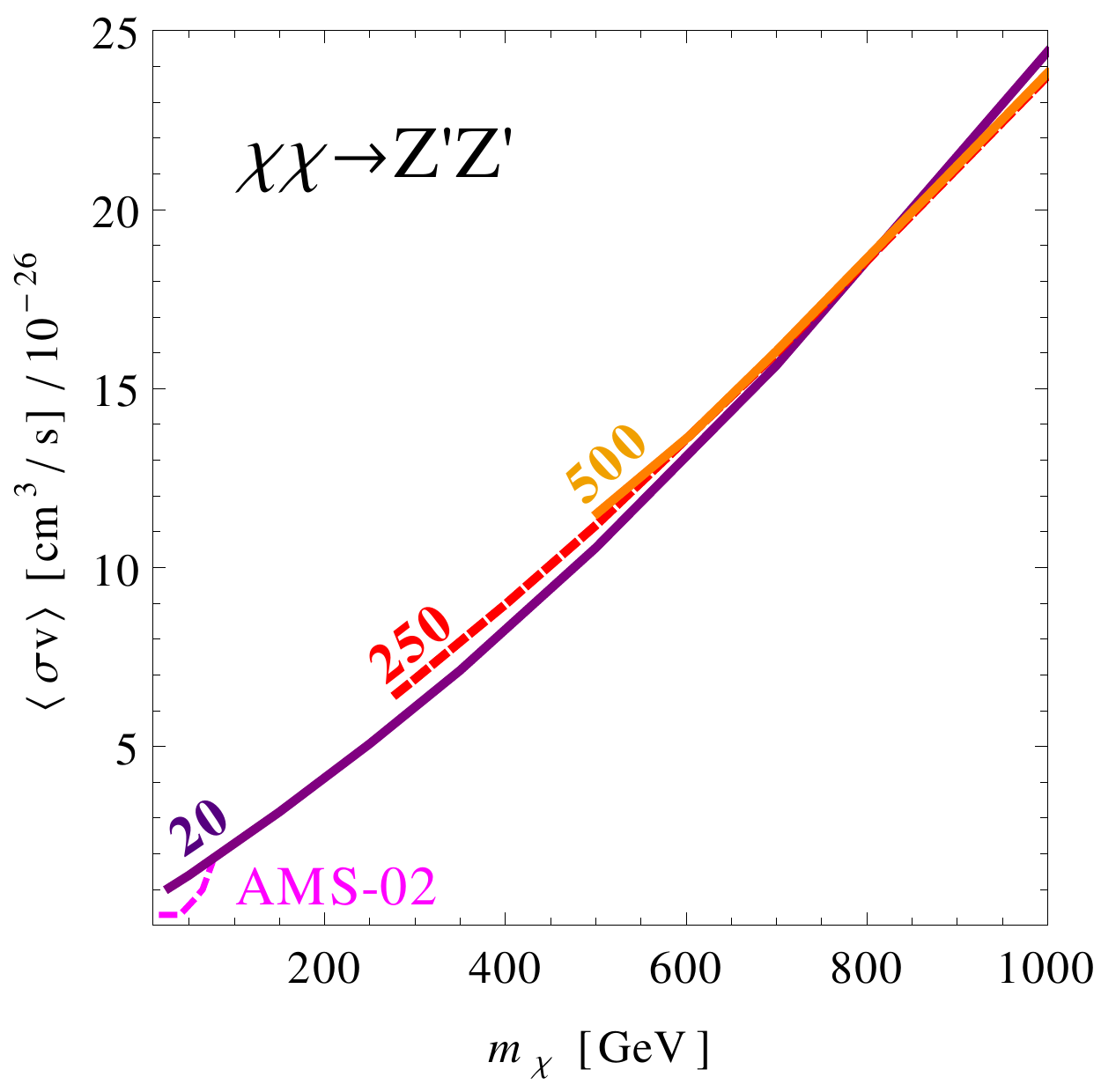}}
\caption{95$\%$ confidence limits (C.L.) on the annihilation cross
  section from Fermi Pass 8 data on 15 dwarf spheroidal galaxies.  Limits on the $sZ'$ process 
  are shown for dark Higgs masses of 20, 100 and 200 GeV, for various $Z'$ masses as labeled in the plots. $Z'$ masses stated are in GeV.
  Limits on the $Z'Z'$ process are shown for $Z'$ masses of 20, 250 and 500 GeV. 
  The approximate limit from AMS-02 is shown as a dashed magenta line, and is only applicable if the sum of the final state mediators is less than about 70 GeV.
  Intermediate mediator mass limits can be simply obtained from interpolation of these plots. All these plots can be applicable to any of the scenarios outlined in this paper: 
  the appropriate limit of $sZ'$ or $Z'Z'$ will depend on the specific choices of the couplings, masses and for which process is kinematically allowed.}
\label{fig:fermix1}
\end{figure}

The spectra generated are then compared to the strongest indirect
detection limits available for our processes\footnote{We also include the approximate limit from AMS-02 at low DM masses, adapted from \cite{Elor:2015bho}. This approximate limit is only applicable if the sum of the final state mediators is less than about 70 GeV.}: the Fermi-LAT Pass~8
data on dwarf spheroidal galaxies (dSphs) of the Milky
Way~\cite{Ackermann:2015zua}.  To find the limit on the cross section
from dSphs, we use the maximal likelihood method to compare our
spectra against those for the dSphs publicly provided by Fermi-LAT in
the Pass 8 data, with the $J$ factor taken to be a nuisance parameter
as per Ref.~\cite{Ackermann:2015zua}.  We take spectra from 15 dSphs:
Bootes I, Canes Venatici II, Carina, Coma Berenices, Draco, Fornax,
Hercules, Leo II, Leo IV, Sculptor, Segue 1, Sextans, Ursa Major II,
Ursa Minor, and Willman 1.  The 95$\%$ C.L. limits on the annihilation
cross section are shown Fig.~(\ref{fig:fermix1}), for various dark
Higgs and $Z'$ masses, for both $sZ'$ and $Z'Z'$ processes.

The limits we show are independently set on either the
$\chi\overline{\chi}\rightarrow sZ'$ process or the
$\chi\overline{\chi}\rightarrow Z'Z'$ process.  They can then be applied
to any of the scenarios we study in this paper, assuming that one of
the modes dominates.  Indeed, they can also be applied to any model
that features annihilations to a $sZ'$ or $Z'Z'$ final state, provided
the $Z'$ and $s$ communicate with the SM via kinetic or Higgs mass
mixing respectively, as the cross section limits depend only on the
gamma-ray spectral shape that characterizes a given annihilation mode.
From Fig.~(\ref{fig:fermix1}) it is clear that the limit on the cross
section does not vary greatly with the mediator mass provided it is
kinematically allowed; it is instead the DM mass with which
the energy of final state photons and thus cross section limits is
tightly correlated.

The thermal relic cross section required to reproduce the correct
relic density for \mbox{non-self} conjugate DM is approximately $\langle\sigma v\rangle\approx 4.4
\times10^{-26}$ cm$^3$/s \cite{Steigman:2012nb}, which excludes the low DM mass region where the
Fermi limits surpass this sensitivity.  However, this statement
assumes that the $s$-wave contributions to the annihilation cross
section dominate both at freeze-out and in the universe today.  In fact
in some cases, such as scenario III, the $p$-wave processes can make a
non-negligible contribution at freeze-out.  This means that the relic
density constraint could be satisfied, yet the cross section in the
universe today suppressed, escaping indirect detection bounds even
for low DM mass.

\section{Discussion and Summary}
\label{sec:summary}

We have surveyed a spectrum of phenomenologically distinct
two-mediator DM models, containing both a dark vector and dark scalar,
where gauge invariance is respected and the mass terms for the
dark sector fields are introduced in a self-consistent way.  These
two-mediator DM models correctly capture important phenomenology which
is missing in the single mediator approach.  Specifically, we modified
the usual simplified model setup to incorporate mass generation for
the DM candidate and vector mediator, by using combinations of bare
mass terms, Higgs mechanisms and Stueckelberg mechansims.  We found
that the DM interaction types and annihilation processes, and hence
both the relic density and indirect detection constraints, are
strongly dictated by the mass generation mechanism we choose for the
dark sector particles:
\begin{itemize}
\item
Unless the DM and $Z'$ masses both receive contributions from the vev
of the same dark Higgs field, pure vector couplings of the spin-1
mediator and DM are required, as discussed in scenarios III
and IV.  In these scenarios DM annihilates to both $sZ'$ and $Z'Z'$,
with the relative rates to these final states controlled by
independent coupling constants.  Moreover, in the high energy limit,
only the $Z'_T$ polarization is produced by these annihilations.
\item
However, if a dark Higgs mechanism gives mass to all the dark sector
fields, as per scenario II, the axial-vector coupling between the
spin-1 mediator and DM must be non-zero. In this scenario,
the $sZ'$ and $Z'Z'$ DM annihilation channels are intrinsically
linked.  Furthermore, production of the $Z'_L$ polarization enhances
the annihilation to $sZ'$.  If both the vector and axial-vector
couplings are non-zero, the annihilation to $Z'Z'$ is also enhanced by
$Z'_L$ (via the $V-A$ interference) though it remains subdominant to
the $sZ'$ mode when both are kinematically allowed.
\end{itemize}

One may imagine generalizations of scenarios III and IV in which the
$Z'$ and $\chi$ masses arise from {\it two different} Higgs
mechanisms.  Indeed, we would recover scenario III (Stueckelberg $Z'$
mass) in the limit that the Higgs responsible for the $Z'$ mass is
taken to infinity.  Likewise, we would recover scenario IV (bare
$\chi$ mass) in the limit that the Higgs responsible for the $\chi$
mass is taken very large.  In these generalizations, the $\chi$-$Z'$
coupling remains of pure vector form.  Axial couplings always imply
that a Higgs which Yukawa couples to the $\chi$ must carry $U(1)_\chi$
charge, and hence its vev also contributes to the $Z'$ mass, as in
scenario II.  Such two-scalar models would lead to additional
complexity via mixing in the scalar sector, but would not introduce
any qualitatively new $Z'$ physics.

Our results are not captured by the single mediator approach, where
the mass generation mechanism is left unspecified and constraints on
the coupling types are not usually applied.  This means that by
continuing to use simplified models with a single spin-1 mediator, (i)
we are at best only testing a very specific subset of the
possibilities: Dirac DM with a bare mass and pure vector couplings to
a $Z'$ with a Stueckelberg derived mass (i.e. scenario I) or (ii) at
worst, experimental constraints may not be meaningful because the
models have been oversimplified.  Option (i) is not particularly
appealing in that it does not cover well motivated possibilities such as
Higgs mass generation (which, after all, is a mechanism we know is
realized by nature) or Majorana DM.  The remaining option, (ii), is
far from desirable.

\section*{Acknowledgements}

This work was supported in part by the Australian Research Council. RKL thanks John Beacom and the Center for Cosmology and AstroParticle Physics (CCAPP) at Ohio State University for their hospitality and support during her visit, where part of this work was completed. Feynman diagrams are drawn using {\sc TikZ-Feynman} \cite{Ellis:2016jkw}. 

\section*{Appendix}
\appendix

\section{Cross Sections}

In the scenarios discussed in this paper, the charges were fixed to
particular values either to satisfy gauge invariance, or to
demonstrate the phenomenology. The full cross sections with explicit
$Q_{A,V}$ and $Q_S$ dependence are listed in this appendix for
reference.

The full $s$-wave cross section for $\chi\overline{\chi}\rightarrow Z'Z'$ is
\begin{align}
 \langle \sigma v \rangle_{\chi\overline{\chi}\to Z'Z'} & = \frac{g_{\chi }^4 \left(1-\eta _{Z'}\right)^{3/2} \left(2 Q_A^2 Q_V^2 \left(4-3 \eta _{Z'}\right)+Q_A^4 \eta _{Z'}+Q_V^4 \eta _{Z'}\right)}{4 \pi  m_{\chi }^2 \left(\eta _{Z'}-2\right)^2 \eta _{Z'}}.
\end{align}
This expression gives the $s$-wave contribution to the
$\chi\overline{\chi}\rightarrow Z'Z'$ cross section for {\it all} cases, as only
the $t/u$ channel diagrams contribute.  (Scalar-mediator contributions
only enter at the $p$-wave level.)  We see that if either $Q_A$ or
$Q_V$ is zero, the cross section scales as $1/m_\chi^2$ in the limit
that $\eta_{Z'}= m^2_{Z'}/m^2_\chi \ll 1$, and is dominated by $Z_T'$
contributions only.  In the case that both $Q_A$ and $Q_V$ are
non-zero, the cross section instead scales as $1/m_{Z'}^2$ in the
$\eta_{Z'} \ll 1$ limit, which arises due to the $Z_L'$ modes.  Note
however, that no violation of unitarity will occur -- the $Z'$ mass
cannot be made arbitrarily large while satisfying the constraint
Eq.~(\ref{eq:coupling}) and restricting all couplings to perturbative
values. This $Z'Z'$ cross section matches that in Refs. \cite{Alves:2015pea,Alves:2015mua}.

The full $s$-wave cross section for $\chi\overline{\chi}\rightarrow sZ'$ is
\begin{align}
 \langle \sigma v \rangle_{\chi\overline{\chi}\to sZ'} & =\frac{g_{\chi }^2 \sqrt{ \left(\eta _s - \eta _{Z'} -4 \right)^2 - 16 \eta_{Z'} }}{256 \pi  m_{\chi }^2 \left(\eta _{Z'}-4\right)^2 \eta _{Z'}^2 \left(\eta _s+\eta _{Z'}-4\right)^2}\times\\
 &\Big\{g_{\chi}^2 Q_S^2 \left(\eta _s+\eta _{Z'}-4\right)^2 \Big[Q_A^2 \left(\eta _{Z'}-4\right){}^2 \left( \left(\eta _s - \eta _{Z'} -4 \right)^2 - 16 \eta_{Z'} \right)\nonumber\\
 &+Q_V^2 \eta _{Z'}^2 \left( \left(\eta _s - \eta _{Z'} -4 \right)^2 +32 \eta_{Z'}  \right)\Big]\nonumber\\
 &+24 \sqrt{2} g_{\chi } Q_S Q_V^2 y_{\chi } \left(\eta _{Z'}-4\right) \eta _{Z'}^{5/2} \left(\eta _s^2-8 \eta _s-\eta _{Z'}^2+16\right)\nonumber\\
 &+4 Q_V^2 y_{\chi }^2 \left(\eta _{Z'}-4\right)^2 \eta _{Z'}^2 \left( \left(\eta _s - \eta _{Z'} -4 \right)^2 +8 \eta_{Z'} \right)\Big\}\nonumber.
\end{align}
Taking $2Q_A= Q_S=1$ and using the relation of the Yukawa and gauge coupling in Eq.~(\ref{eq:coupling}) recovers the cross sections for scenario II, $Q_A=Q_S= 0$ recovers the cross sections for scenario III, and $Q_A, y_\chi= 0$ gives the cross sections for scenario IV.

It is still important to note however that the values for the charges
cannot be chosen freely and should obey the constraints discussed in
this paper.

\bibliographystyle{JHEP}
\bibliography{darkhiggs_bib}

\end{document}